# Are there universal signatures of topological phases in high harmonic generation? Probably not.

Ofer Neufeld[1,*], Nicolas Tancogne-Dejean[1], Hannes Hübener[1], Umberto De Giovannini[1,2], and Angel Rubio[1,3,*]

[1]Max Planck Institute for the Structure and Dynamics of Matter and Center for Free-Electron Laser Science, Hamburg, Germany, 22761.
[2]Università degli Studi di Palermo, Dipartimento di Fisica e Chimica—Emilio Segrè, Palermo I-90123, Italy.
[3]Center for Computational Quantum Physics (CCQ), The Flatiron Institute, New York, NY, USA, 10010.
*Corresponding author E-mails: ofer.neufeld@gmail.com, angel.rubio@mpsd.mpg.de

High harmonic generation (HHG) has developed in recent years as a promising tool for ultrafast materials spectroscopy. At the forefront of these advancements, several works proposed to use HHG as an all-optical probe for topology of quantum matter by identifying its signatures in the emission spectra. However, it remains unclear if such spectral signatures are indeed a robust and general approach for probing topology. To address this point, we perform here a fully *ab-initio* study of HHG from prototypical two-dimensional topological insulators in the Kane-Mele quantum spin-Hall and anomalous-Hall phases. We analyze the spectra and previously proposed topological signatures by comparing HHG from the topological and trivial phases. We demonstrate and provide detailed microscopic explanations of why in these systems none of the thus far proposed observables uniquely and universally probe material topology. Specifically, we find that the: (i) HHG helicity, (ii) anomalous HHG ellipticity, (iii) HHG elliptical dichroism and (iv) temporal delays in HHG emission, are all unreliable signatures of topological phases. Our results suggest that extreme care must be taken when interpreting HHG spectra for topological signatures, and that contributions from the crystal symmetries and chemical nature might be dominant over those from topology. They hint that a truly universal topological signature in nonlinear optics is unlikely, and raise important questions regarding possible utilization and detection of topology in systems out-of-equilibrium.

## I. INTRODUCTION

High harmonic generation (HHG) is a non-perturbative nonlinear optical process that occurs when intense laser fields interact with matter, causing the emission of high energy photons [1–3]. This process was discovered in the gas phase, and has been extensively applied to condensed phases of matter over the last decade [3–10]. One of the great promises of HHG from solids that has been pushing the field, is its ability to provide accurate time-resolved data about the driven material properties, including ultrafast electron dynamics [11–14], band structures and Berry curvatures [15–18], symmetries [19–26], electron density information [27], information about light-driven phases of matter [27–31], ultrafast correlations [30,32], phonon dynamics [33–35], etc. Since high harmonic spectroscopy (HHS) is inherently all-optical, table-top, and can theoretically reach a temporal resolution of few hundreds of attoseconds, it is a very appealing technique. On the other hand, it is often limited by the need to extensively compare experimental data with theoretical simulations in order to pinpoint the exact effects and physical interpretation. The fact that the underlying microscopic mechanism and physical picture for HHG in solids remains somewhat controversial and not fully understood [36–39] makes such analyses even more difficult, especially for simplified models that employ approximations (e.g. neglecting electron-electron interactions, utilizing only few active bands, employing tight-binding approximations, employing low-dimensionality models, neglecting spin-orbit interactions, etc.). Moreover, it remains unclear if the strong laser field changes the electronic properties of matter in a way that affects their potential observation in HHG spectra (e.g. *via* Floquet physics [29–31]). Overall, there is a vital and growing need for unbiased *ab-initio* predictions of HHG from quantum materials to establish which properties can or cannot be extracted from this nonlinear spectroscopy tool.

One potentially appealing application of HHS is to extract information about material topology. In this framework, one should drive and measure HHG from materials with non-trivial topological bands, such as topological insulators (TI) [40], Dirac or Weyl semimetals [41,42], quantum-Hall systems, or other states of matter [43–46]. In practical terms, such topological phases can arise when the Berry curvature integrated over the full Brillouin zone (BZ) for a given band is nonzero, leading to nonzero Chern numbers. The hope for HHS is that the Chern numbers and/or other topological invariants are: (i) Imprinted onto the HHG spectrum, and (ii) are imprinted in a manner that allows their extraction in post-analysis. Notably, measuring such topological invariants is an extremely challenging task that is often only accomplished by a combination of optical and electronic techniques such as angle-resolved photoelectron spectroscopy [47–51], low-temperature transport measurements [52], and nonlinear currents [53], and is an active area of research [54–60].

Using HHS, it was predicted that harmonic emission from TI should be elliptically polarized even when driven by linearly polarized light along high-symmetry axes [61–63]. The helicity of the HHG emission was predicted to flip sign in the Haldane model at the phase transition, thus imprinting topological phase information onto the HHG spectra. Alternatively, several works utilized HHG measurements *vs.* the driving ellipticity and proposed topological features in the HHG yield when it maximizes for nonzero or circular ellipticities [64–66], and in its elliptical dichroic response [66]. In Ref. [66] for instance, by comparing to theoretical simulations, the topological contribution to the 'anomalous' (i.e., non-atomic-like) HHG ellipticity dependence could be traced to the presence of topological surface states (while the bulk did not exhibit any anomalous behavior), and an elliptical dichroic response was reported to be much stronger in the topological phase. There have also been other studies reporting possible HHG signatures of topology, including enhanced carrier-envelope-phase sensitivity [67] (which was recently argued to arise from chirped driving [68]), symmetry-forbidden harmonics [69] (which have been established to originate from broken inversion symmetry at the surface [64]), resonant harmonic peaks [70], half-cycle temporal delays in the HHG emission [71], and HHG yield enhancements [72]. In all of these examples, the theoretical predictions were at the heart of the interpretation process that allows extracting spectroscopic information, making it even more crucial to obtain *ab-initio* predictions for HHG from topological systems. Notably, all of the studies above employed some degree of approximations, either using model Hamiltonians by design, or by neglecting potentially important terms.

Here we perform *ab-initio* simulations of HHG in quasi-two-dimensional (2D) TI with state-of-the-art time-dependent spin density functional theory (TDSDFT). Our approach naturally incorporates the system's complete band and crystal structures, electron-electron interactions, spin-orbit coupling (SOC), and electron-ion interactions, which allows drawing conclusions for phenomena in realistic systems [73]. In this work we focus on the HHG emission from bismuthumane (BiH) [74], a material exemplary of the quantum spin-Hall TI phase [75], and a monolayer of $Na_3Bi$, which is exemplary of the quantum anomalous-Hall TI phase [76]. We explore the HHG dependence (yield, polarization, and temporal characteristics) in the driving ellipticity and angle, and test possible topological signatures proposed in the literature. These results are compared to those from the topologically-trivial phases with SOC switched off. From this extensive numerical study, we draw several main conclusions: (i) The HHG helicity generally does not map the topological phase diagram. (ii) An anomalous ellipticity dependence for the HHG yield is not necessarily connected to topological surface states. (iii) HHG elliptical dichroism can arise in both trivial and topological phases with a similar magnitude, and is largely a result of crystal structure (rather than carrying unique topological information). (iv) Temporal delays in HHG emission are a poor indicator of the topological phase. Our results stress that caution must be taken when analyzing HHG spectra for topological fingerprints, and when attempting to extract the material topological nature from measurements.

## II. METHODOLOGY

We begin by outlining our methodological approach and chosen material systems. Calculations for an exemplary quantum spin-Hall TI were performed in the 2D bismuthumane (BiH). BiH exhibits a honeycomb lattice where each bismuth atom is bonded to an additional hydrogen atom in a staggered configuration that preserves inversion symmetry [74] (see Fig. 1(a)). We further note that the lattice structure exhibits three mirror planes transverse to the monolayer, three 2-fold rotational axes that are transverse to the bismuth-bismuth bonds, a 3-fold rotational axis transverse to the monolayer plane, and a 6-fold improper-rotational axis transverse to the monolayer plane (see Fig. 1(a)). In the absence of SOC, BiH is a topologically-trivial semimetal with linearly-dispersing Dirac bands with degenerate band touching points at K and K', similar to graphene (see Fig. 1(b)). With SOC, it enters through a Kane-Mele mechanism into a TI phase with an indirect topological gap of ~1eV, and a minimal direct gap of ~1.35eV at the K and K' points (within the local spin density approximation (LSDA)). This system preserves inversion and time-reversal symmetries such that each state is spin-degenerate, with spin-degenerate bands carrying opposite Berry curvatures and Chern numbers). It thus inherently differs from the Haldane model [61–63].

Calculations for an exemplary quantum anomalous-Hall TI were performed in a monolayer of $Na_3Bi$ [76], which exhibits a hexagonal lattice with broken inversion symmetry, and three-fold rotational symmetry along with three mirror planes (see Fig. 1(c,d)). In the absence of SOC $Na_3Bi$ is a topologically trivial insulator with a direct gap of ~0.1eV at the Γ-point. SOC induces band inversion, creating a direct topological gap of ~0.4eV at Γ (within the LSDA). Since each spin channel supports distinct topological bands in 2D, this system is a physically close realization to the TI phase in the Haldane model.

The ground states (GS) are obtained with non-collinear spin density functional theory (SDFT) [77], with the LSDA for the exchange-correlation (XC) functional, and incorporating semi-relativistic corrections including SOC. The Kohn-Sham-Bloch (KSB) states are obtained by employing periodic boundary conditions in the two in-plane dimensions (*xy* plane) with a discretized *Γ*-centered *k*-grid for sampling the 2D BZ, while allowing a non-periodic *z*-axis with additional vacuum spacing above and below the monolayer. All calculations are performed with the open access real-space real-time code Octopus [78–80]. Further technical details are delegated to Appendix A.

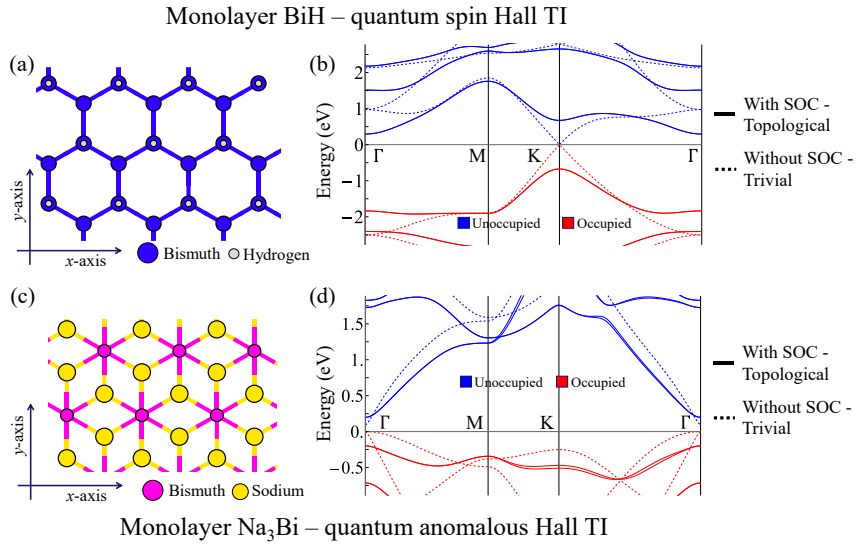

FIG. 1. Geometry and band structure of monolayers of BiH and $Na_3Bi$. (a) Illustration of the hexagonal BiH lattice. (c) Band structure of BiH with and without SOC, where for the SOC case each band is twice degenerate. Zero energy is set to the highest occupied level in the SOC-free system. (c,d) Same as (a,b), but for $Na_3Bi$.

The interaction of the system with an intense classical laser pulse is modeled within TDSDFT [77] in the velocity gauge and dipole approximation. We neglect ionic motion and assume frozen nuclei. The KSB states are self-consistently propagated starting from the GS initial configuration at $t = 0$ with the following time-dependent equations of motion (given in atomic units, used throughout):

$$i\partial_t |\psi_{n,\mathbf{k}}^{KS}(t)\rangle = \left(\frac{1}{2}\left(-i\mathbf{\nabla} + \frac{\mathbf{A}(t)}{c}\right)^2 \sigma_0 + v_{KS}(t)\right)|\psi_{n,\mathbf{k}}^{KS}(t)\rangle \quad (1)$$

where $|\psi_{n,\mathbf{k}}^{KS}(t)\rangle$ is the KSB state at *k*-point **k** and band index *n*, which is a Pauli spinor:

$$|\psi_{n,\mathbf{k}}^{KS}(t)\rangle = \begin{bmatrix} |\varphi_{n,\mathbf{k},\uparrow}^{KS}(t)\rangle \\ |\varphi_{n,\mathbf{k},\downarrow}^{KS}(t)\rangle \end{bmatrix} \quad (2)$$

with $|\varphi_{n,\mathbf{k},\alpha}^{KS}(t)\rangle$ the spin-up/down part of the spinor states with spin index $\alpha$. $\sigma_0$ in Eq. (1) is a 2×2 identity matrix, $\mathbf{A}(t)$ is the vector potential of the driving laser pulse such that $-\partial_t \mathbf{A}(t) = c\mathbf{E}(t)$ (with $c$ the speed of light), and $v_{KS}(t)$ is the time-dependent KS potential given by:

$$v_{KS}(\mathbf{r},t) = \int d^3r' \frac{n(\mathbf{r}',t)}{|\mathbf{r}-\mathbf{r}'|}\sigma_0 + v_{XC}[\rho(\mathbf{r},t)] + V_{ion} \quad (3)$$

where the first term in Eq. (3) is the classical Hartree term (representing the mean-field Coulombic interaction between electrons) with $n(\mathbf{r},t) = \sum_{n,\mathbf{k},\alpha} w_\mathbf{k} |\langle \mathbf{r}|\varphi_{n,\mathbf{k},\alpha}^{KS}(t)\rangle|^2$ the time-dependent electron density, $w_\mathbf{k}$ the *k*-point weights, and the sum runs over all occupied bands. $v_{XC}$ is the XC potential, which is a functional of the time-dependent spin density matrix:

$$\rho(\mathbf{r},t) = \frac{1}{2}n(\mathbf{r},t)\sigma_0 + \frac{1}{2}\mathbf{m}(\mathbf{r},t)\cdot\boldsymbol{\sigma} \quad (4)$$

where $\boldsymbol{\sigma}$ is a vector of Pauli matrices and $\mathbf{m}(\mathbf{r},t)$ is the time-dependent magnetization vector:

$$\mathbf{m}(\mathbf{r},t) = \sum_{n,\mathbf{k}} w_\mathbf{k} \langle \psi_{n,\mathbf{k}}^{KS}(t)|\mathbf{r}\rangle \boldsymbol{\sigma} \langle \mathbf{r}|\psi_{n,\mathbf{k}}^{KS}(t)\rangle \quad (5)$$

$V_{ion}$ in Eq. (3) represents the interactions of electrons with the lattice ions and core electrons (through a nonlocal pseudopotential), and incorporates scalar relativistic correction terms as well as the spin-orbit coupling term, resulting in non-diagonal contributions in spin-space that couple the different spin channels. Note that that the temporal dependence of $v_{KS}(\mathbf{r},t)$ (which arises as the electron density evolves in time when the system interacts with the laser field) effectively couples the different band and *k*-point indices of the KSB states and their equations of motion through electron-electron interactions (including both classical Coulombic repulsion and quantum-mechanical effects incorporated in the XC functional). For the temporal propagation an additional complex absorbing potential (CAP) is added along the edges of the *z*-axis to avoid spurious reflection of electrons. All other technical details of the numerical procedures are found in Appendix A.

The driving dipolar laser field is modeled with the following elliptically-polarized vector potential:

$$\mathbf{A}(t) = f(t) \frac{cE_0}{\omega} \frac{1}{\sqrt{1+\varepsilon^2}} \hat{R}(\theta) \cdot [cos(\omega t)\hat{\mathbf{x}} + \varepsilon sin(\omega t)\hat{\mathbf{y}}] \quad (6)$$

where $f(t)$ is an envelope function (see the Appendix A for details), $E_0$ is the field amplitude, ω is the carrier frequency, $\varepsilon$ is the laser ellipticity, and $\hat{R}(\theta)$ is a 2D rotation matrix acting on the polarization degrees of freedom which rotates the main elliptical laser axis in the *xy* plane (such that for $\theta = 0$ the laser is polarized along the *x*-axis, and for $\theta = 90°$ it is polarized along the *y*-axis (see Fig. 1 for orientation schematics).

To calculate the HHG spectra we first obtain the total electronic current in the system, $\mathbf{J}(t) = \int_\Omega d^3r \mathbf{j}(\mathbf{r},t)$, which is obtained by integrating the microscopic time-dependent electronic current density, $\mathbf{j}(\mathbf{r},t)$, over the unit cell, with $\Omega$ the unit cell volume, and with $\mathbf{j}(\mathbf{r},t)$ obtained by integrating the current operator over the full BZ:

$$\mathbf{j}(\mathbf{r},t) = \sum_{n,\mathbf{k},\alpha} \left[ w_\mathbf{k} \varphi_{n,\mathbf{k},\alpha}^{*KS}(\mathbf{r},t) \left( \frac{1}{2}\left(-i\boldsymbol{\nabla} + \frac{\mathbf{A}(t)}{c}\right) + [V_{ion},\mathbf{r}] \right) \varphi_{n,\mathbf{k},\alpha}^{KS}(\mathbf{r},t) + c.c \right] + \mathbf{j_m}(\mathbf{r},t) \quad (7)$$

, where $\mathbf{j_m}(\mathbf{r},t)$ in Eq. (7) is the magnetization current density that vanishes after spatial integration. The HHG spectra is calculated from $\mathbf{J}(t)$ by a discrete Fourier transform after a temporal derivative and multiplying by a temporal window similar to the pulse envelope:

$$I(\Omega) = \left|\int f(t)\, \partial_t \mathbf{J}(t) e^{-i\Omega t} dt\right|^2 \quad (8)$$

Note that Eq. (8) is evaluated only for the *xy*-polarized components of the HHG emission, because *z*-polarized emission is not phase matched and does not propagate to the far-field. From Eq. (8) we also evaluate the ellipticity-helicity of the emitted harmonics through calculations of the Stokes parameters, and separate emission contributions that are polarized along/transversely to the main driving angle.

Importantly, all numerical calculations are performed twice – once for the system with SOC, and once without SOC. For calculations where the SOC term is switched off the XC functional is taken as the standard LDA with unpolarized KSB states (each state is twice degenerate and occupied by two electrons).

## III. RESULTS
### A. Linearly-polarized driving – Parallel and transverse responses

We begin by exploring the nonlinear optical responses of both materials and phases to intense linearly-polarized pulses. Before addressing the deviations in the HHG response between the topological and trivial phases, let us first analyze their similarities and general properties that are undoubtably independent of the topology. In Appendix B and D we present the full data set of HHG spectra *vs.* driving angle in both materials and phases (both unprocessed spectra in Appendix B Figs. 7, 8, and the integrated yields per harmonic order in Appendix D Figs. 10-13). From these spectra we observe several characteristic features. First, the HHG emission from BiH in both phases exhibits sharp odd-only harmonic peaks with a first dominant plateau ranging up to about the 21$^{st}$ harmonic order (~9eV), and a weaker second plateau ranging up to about the 31$^{st}$ harmonic (~13eV). The odd-only harmonic structure is attributed to the dynamical inversion symmetry of the lattice and the monochromatic laser [25]. One might initially wonder if such a symmetry should be broken by the presence of spin-orbit interactions in the topological phase. To better understand the observed selection rule, we analyze the symmetries of the SOC term under inversion. In periodic solids, spin-orbit interactions take the form: $H_{so} = \frac{-i}{4c^2}(\boldsymbol{\nabla}V(\mathbf{r}) \times \boldsymbol{\nabla})\cdot\boldsymbol{\sigma}$, with $V(\mathbf{r})$ representing the mean-field potential felt by electrons in the lattice (equivalent to $v_{KS}$ above) [81]. If $V(\mathbf{r})$ is inversion-symmetric (as is the case for BiH), $\boldsymbol{\nabla}V(\mathbf{r})$ is inversion-odd; and since $\boldsymbol{\nabla}$ is also inversion-odd, $H_{so}$ is overall inversion-symmetric. Similar analysis holds for any of the point-group symmetries of the lattice, such that SOC never induces a symmetry breaking (it also preserves time-reversal and time-translation symmetries crucial for dynamical symmetries [25]). Consequently, any nonlinear optical selection rule that arises from the full electronic current applies to both the system with, and without, SOC. On the other hand, it is important to note that with SOC the responses from each individual spin channel (i.e. the spin currents) do not necessarily respect the symmetries of the Hamiltonian, and thus might not uphold the expected selection rules. In other words, the characteristic response cannot be described by a single spin component state (this is analogous to symmetry-breaking in atomic systems where the response from $p_+$ orbitals might break a certain symmetry, but the full ensemble averaged response does not [82]). Nevertheless, after summing all occupied states and spin channels, and assuming that the system was initially in its ground state, the selection rules are upheld (see further discussion in Appendix C). We recently showed that a similar notion of symmetry affected magnetization dynamics in SOC systems [83].

Second, we note that the HHG emission from monolayer $Na_3Bi$ in both phases similarly respects material symmetries, but here even harmonics are also emitted due to absence of inversion symmetry (see Fig. 2 and Appendix B, D for full data set). We found that the even harmonics are generally weaker than the odd harmonic emission, and arise in a roughly similar magnitude in both the trivial and topological phase, such that their presence does not provide information about any topological character.

Third, by analyzing the polarization-resolved HHG yield (presented in Fig. 2 and Appendix D, Figs. 11, 13), we found that the parallel emission components form the dominant HHG response – the transverse components account for only ~10% of the yield in both material systems. Moreover, the transverse emission has a largely similar magnitude in both the topological and trivial phases (with slight variations depending on the harmonic order). This points towards a dominant role of the lattice structure in inducing the transverse response, and not so much of the SOC or Berry curvature terms [84].

Fourth, the transverse HHG emission in both materials and phases complies with the fundamental laser-matter system's symmetries: in BiH it vanishes when driving along the high-symmetry axes. For $\theta=0°,60°$, this arises from 2-fold rotational axes [25]. For $\theta=30°,90°$ in both $Na_3Bi$ and BiH it vanishes due to mirror planes [25,82,85]. Note again that the HHG emission from each individual spin channel does not respect these symmetries and emits a transverse component (see Appendix C). However, as expected, in BiH both spin channels have opposite signs for the transverse response in the quantum spin-Hall phase, and they vanish after summation (which corresponds to the full observable electronic current in HHG). In $Na_3Bi$ the two spin channels have a chiral-like emission profile that differs, and the overall transverse current does not vanish (as in the driven Haldane model).

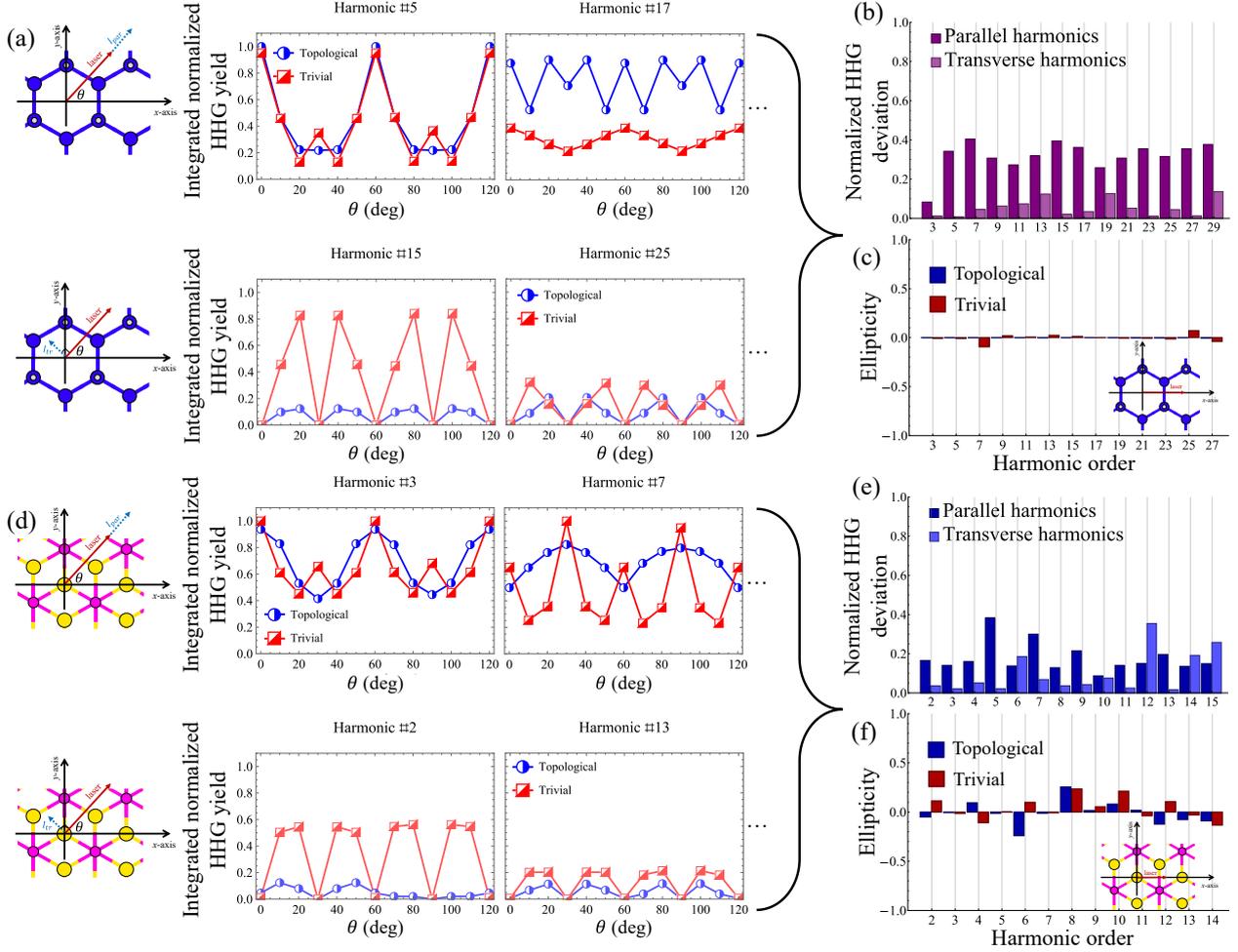

FIG. 2. Parallel and transverse HHG yield *vs.* crystal orientation from monolayers of BiH and Na$_3$Bi driven by a linearly-polarized laser, and resulting deviation between the topological phases in yield and ellipticity. (a) Emission from both topologically-trivial BiH without SOC (red) and the TI phase (blue), for HHG components parallel to the driving axis (top), and transverse (bottom). Only two exemplary harmonic orders are shown (for full data set see Appendix D, Figs. 10-13). The yields are normalized to the peak HHG emission for each harmonic order. (b) Normalized deviation between the topological phases in the angle-integrated HHG yield, resolved for each polarization component. (c) HHG ellipticity per harmonic order. The emission is linearly-polarized in BiH (zero ellipticity) due to material symmetries. The ellipticity per harmonic order is calculated by weighting the spectrally-resolved ellipticity with the HHG intensity for *x*-axis driving. (d,e,f) Same as (a,b,c) but for Na$_3$Bi material system, and where the emission ellipticity is nonzero in (f). Plots calculated for a driving wavelength of 3000nm and peak intensity of $2\times10^{11}$ W/cm$^2$ for BiH and $10^{11}$ W/cm$^2$ for Na$_3$Bi.

At this stage, let us analyze potential deviations and topological signatures in the HHG spectra in both materials. Figures 2(c,f) present the harmonic-order-resolved HHG ellipticity in both materials and phases upon driving along the *x*-axis. Most importantly, our *ab-initio* results highlight that: (i) In BiH, since a transverse response is absent along the high-symmetry axes, harmonics are emitted with linear polarization rather than elliptical (Fig. 2(c)). This result differs from that obtained in previous works predicting elliptical harmonics from TIs with a helicity that flips sign at every topological phase transition, which is a result of inherent symmetry breaking in the Haldane model [64–66]. Notwithstanding, elliptical harmonics are indeed emitted when driving along non-high-symmetry axes, but we stress that this is not a topological feature and arises from the crystal structure itself (similar features were also observed from non-topological materials [86]). In our analysis, we could not find a substantial and robust deviation between ellipticities or helicities of harmonics emitted by both phases in this regime (not presented). (ii) In Na$_3$Bi, where a transverse response does not vanish, the harmonic ellipticity is still not a good indicator of a phase transition (Fig. 2(f)). That is, some harmonic orders flip helicity between phases (e.g. the 2$^{nd}$ or 4$^{th}$ harmonics), and some do not (e.g. the 8$^{th}$, 9$^{th}$, 10$^{th}$ harmonics), and there is no sequence rule that determines which harmonics change sign. Overall, there is no consistent flip in the harmonic ellipticity across the phase transition as predicted in the Haldane model [64–66]. Physically, this arises simply because the two phases have distinct band structures (which probably carries a much stronger contribution to the harmonic ellipticity than topology), and because the mechanism that opens a topological gap in the Haldane model differs from that in Na$_3$Bi. Moreover, we note that the HHG ellipticity is in general quite weak and on the order of ~0.1, such that it would be difficult to detect.

One potentially important feature in the HHG spectra, is that it shows a very strong angular dependence that can be strikingly different in the topological and trivial phases. Interestingly, we found that the transverse HHG components (Fig. 2(a,d) bottom rows) display a very similar angular dependence for the two phases. In that respect, this component seems to not strongly depend on the topological nature nor the band structure, despite it carrying anomalous contributions from Berry curvature [87]. We can quantitatively evaluate the degree to which the angular dispersions in the trivial and topological phases differ by defining a normalized deviation between the two that takes into account the shape and phase of the HHG yield *vs.* driving angle (see Appendix A for details). Figure 2(b,e) presents the normalized deviation, showing that it is indeed very small for the transverse components in both materials. Practically, the insensitivity of the transverse emission to topology comes about because this component is exceedingly sensitive to crystal symmetries.

The angular dependence of the parallel HHG component on the other hand, can vary greatly between the phases (Fig. 2(a,d) top rows). Figures 2(b,e) show the normalized deviations that reach ~40% in BiH, and ~20% in Na$_3$Bi. For instance, the 7$^{th}$ harmonic in the topological phase in BiH shows a very strong angular dependence, while in the trivial phase it is almost independent of $\theta$. In some other cases (e.g. harmonics 13, 15, 21, 23, see Appendix D Fig. 11) the yield can maximize at a given angle for one phase, but at another angle for the other. Since the lattice structure in both phases is identical, this result might hint that the HHG spectrum contains potentially useful information for spectroscopy of topology, as has been suggested in previous works. It would be especially appealing here, because the information is embedded in the intense parallel component of the emission. However, we stress that care must be taken before interpreting this effect as a unique fingerprint of material topology. Four main factors prevent a direct connection: (i) The band structure of the two phases is substantially different (see Fig. 1(b,d)), and plays a dominant role in determining the nonlinear optical response. Indeed, the normalized deviation is bigger in BiH than in Na$_3$Bi, probably because there the band structures differ more greatly between the trivial and topological phases. (ii) SOC also affects the electron dynamics [84]. (iii) One can obtain similar angular distributions also in other systems, such that one measurement does not provide a unique signature. For instance, 6-fold structure for the HHG spectra may arise due to a 6-fold rotational symmetry (e.g. as in graphene), a 6-fold improper rotational symmetry (as in BiH), or even a 3-fold rotational symmetry (e.g. as in hexagonal-boron-nitride [34] or MoS$_2$ [88]), and one cannot make a direct connection without performing additional measurements. Moreover, the angular dispersion can be highly sensitive to the laser conditions, hampering their utilization for extracting system-specific information. (iv) We could not find any straightforward way to extract topological information from the data. The only currently available approach is to thoroughly compare measurements with simulations in various phases, but that already implies *a-priori* knowledge about topology, and it's also not always possible to exclude other effects. Moreover, no clear prescription exists for cases where simulations and measurements disagree, which could arise due to a multitude of reasons including macroscopic or microscopic effects not included in the simulations (e.g. phase-matching, re-absorption, phonons, scattering, beyond-dipole contributions, de-coherence, etc.), measurement errors, material defects, surface contribution, and more.

Overall, while our results show some promise for HHG spectroscopy of topology using linearly-polarized driving, further research is required to formulate more elaborate, quantitative, and robust, approaches for extracting such data from spectra. We emphasize that at this stage, we cannot determine to what extent angular-dependent HHG spectra contain contributions from the material topology, rather than being dominated by standard contributions to the dynamics from the band structure, spin-orbit coupling, and Berry curvature, which are generically independent of the topology (in the sense that nonzero Berry curvature appears also in non-topological materials where it still influences HHG). Most importantly, it remains unclear how such contributions could be extracted effectively, even if they exist.

### B. Elliptically-polarized driving - Anomalous ellipticity

Next, we explore the HHG yield under elliptically-polarized driving. For this purpose, we first fix the main elliptical axis to either the *x* or *y* axes and vary the ellipticity, $\varepsilon$. Figure 3 shows the resulting integrated HHG yield per harmonic order *vs.* $\varepsilon$ in both systems (see Appendix D, Figs. 14-16 for full data sets). In BiH, we identify odd-only harmonic peaks throughout the spectra for any value of $\varepsilon$ due to the inversion symmetry, while for circularly-polarized driving only *6n±1* harmonic orders are allowed (for integer *n*). The appearance of only *6n±1* harmonic orders for circularly-polarized driving (even though BiH has a 3-fold rotational symmetry that usually leads to *3n±1* harmonic selection rules) is a result of the 6-fold improper-rotational symmetry axis in BiH.

Notably, the spectra show a combination of 'regular' HHG behavior (where the HHG yield reduces with ellipticity), and 'anomalous' (where the yield maximizes for elliptical driving). For a three-dimensional (3D) TI, it was recently suggested that anomalous ellipticity dependence could be a telltale sign of topology because it was only exhibited by the topological phase, and reconstructed in model calculations from the topological surface states [66]. Our results show that in the 2D case (both spin-Hall and anomalous-Hall) this is not necessarily an effective approach as both phases provide anomalous contributions in the bulk. Moreover, the functional behavior of the HHG yield strongly depends on the driving axis. For instance, harmonic #5 in the trivial phase of BiH behaves anomalously when driven along the *y*-axis, but regularly when driven along the *x*-axis. It also depends on the material system – when driven along the *x*-axis, the 5$^{th}$ harmonic shows anomalous behavior in BiH only for the trivial phase, but in Na$_3$Bi the behavior flips and an anomalous behavior is observed only for the topological phase. Additionally, harmonics can behave similarly in both phases and carry no direct signature (e.g. harmonic #11 in Fig. 3(a)). Such similar behavior in HHG responses in both phases is generally more prominent in Na$_3$Bi (see also Appendix D, Fig. 16), where the band structure differences between phases are relatively small. This again hints that the band structure plays a dominant role in determining the HHG characteristics. Overall, we conclude that the mere existence (or absence) of an anomalous ellipticity dependence in the HHG yield is not universally or uniquely connected with any topological feature.

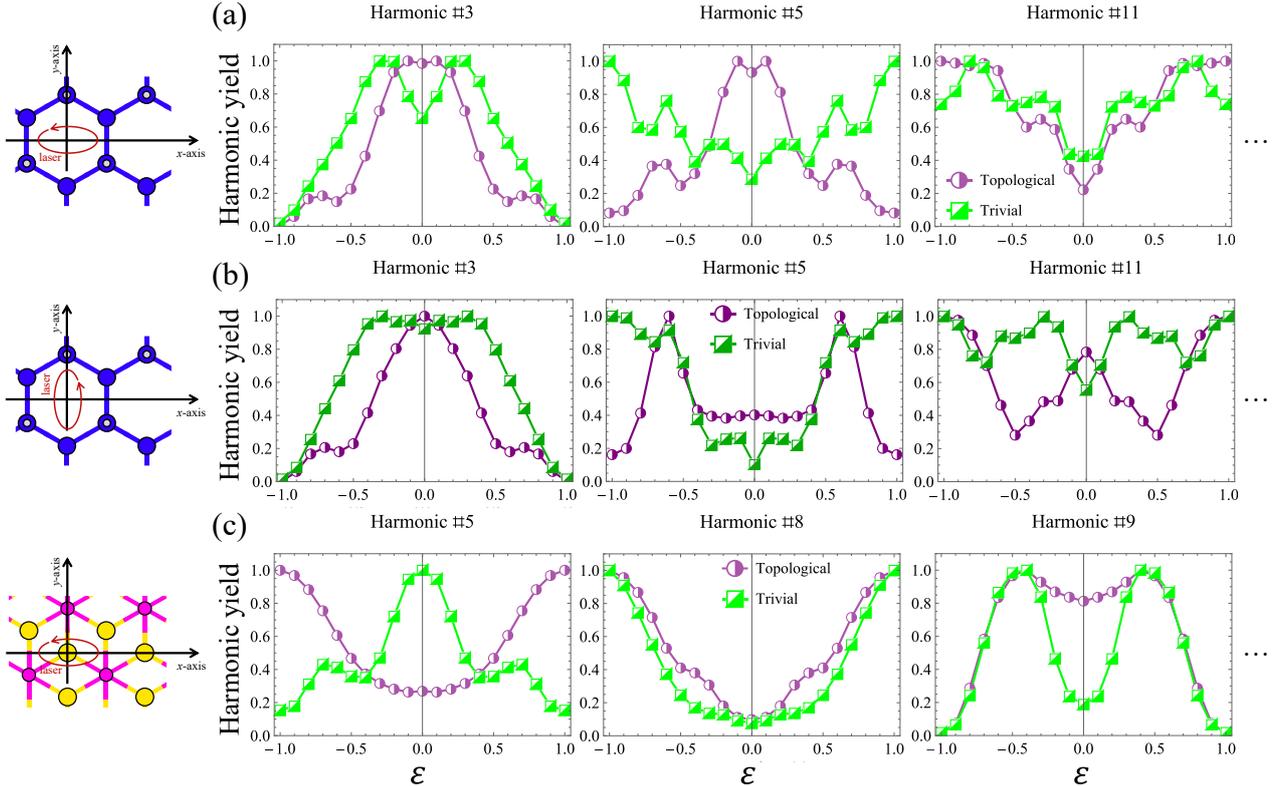

FIG. 3. HHG yield per harmonic order *vs.* driving ellipticity from monolayers of BiH and Na$_3$Bi. (a) HHG yield for exemplary harmonics in BiH for trivial (green) and topological (purple) phases (for full data set see Appendix D, Figs. 14-16), where the elliptical major axis is oriented along the *x*-axis. (b) Same as (a) but with an elliptical major axis oriented along the *y*-axis. (c) Same as (a) but for Na$_3$Bi. Plot calculated for similar conditions to Fig. 2. Each plot is normalized to the peak HHG yield for each harmonic order.

### C. Elliptically-polarized driving – Elliptical dichroism

We next study HHG elliptical dichroism. Heide *et al.* recently identified such dichroism for a 3D-TI system as a potential source for topological HHS [66]. Note that the elliptical dichroism inherently vanishes when the elliptical major axis points to a high-symmetry axis. We start by exploring this effect in BiH, and perform line scans for fixed $\theta$ away from the high-symmetry *x*-axis while varying $\varepsilon$ from -1 to 1. The dichroism is defined as:

$$ED^n(\varepsilon) = 2\frac{I_\theta^n(\varepsilon) - I_{-\theta}^n(\varepsilon)}{I_\theta^n(\varepsilon) + I_{-\theta}^n(\varepsilon)} \qquad (9)$$

where $I_\theta^n(\varepsilon)$ is the integrated harmonic yield for the *n*'th harmonic driven by an elliptically polarized field with a major axis oriented at $\theta$, with an ellipticity $\varepsilon$, and $ED^n(\varepsilon)$ is the resulting dichroism for the *n*'th harmonic and ellipticity $\varepsilon$, which is bound from -200% to 200%. Alternatively, one could replace Eq. (9) with a more natural definition where the harmonic helicity is inverted:

$$ED^n(\varepsilon) = 2\frac{I_\theta^n(\varepsilon) - I_\theta^n(-\varepsilon)}{I_\theta^n(\varepsilon) + I_\theta^n(-\varepsilon)} \qquad (10)$$

In fact, for BiH (as well as the material probed in ref. [66]), the nonlinear optical response respects the symmetry $\varepsilon \rightarrow -\varepsilon$, $\theta \rightarrow -\theta$, rendering Eqs. (9) and (10) equivalent. This is a result of the mirror symmetry about $\theta=0°$ here, but can also arise due to other symmetries. It's important to note that due to inversion symmetry the system does not exhibit circular dichroism, and in this case we have that $ED^n(\varepsilon = \pm 1) \equiv 0$. A nonzero circular dichroism would indicate the absence of mirror symmetries in the driven crystal plane (e.g. as in quartz [26] or chiral Weyl semimetals [89]).

Figure 4 presents our main results for exemplary harmonics for the benchmark case of $\theta=\pm 15°$ (see Appendix D, Figs. 17, 18, 21, 22 for full data). BiH and Na$_3$Bi both exhibit strong elliptical dichroism in the trivial and topological phases ranging up to ~75%. Unlike the 3D-TI case [66], we could not identify a substantial difference in the robustness or magnitude of the elliptical dichroism between the phases. One potentially promising feature is that the finer structure of the dichroism can vary with the topological phase. Nevertheless, it's unclear if such data is useful for HHS of topology, since there is no obvious way to deconvolve any potential topological contributions from the standard contributions to HHG in both phases. Overall, we conclude that the main contributor to the presence or absence of elliptical dichroism in the solid HHG response is the crystal structure itself rather than topology (at least in these cases).

At this point we wish to discuss a main feature in Fig. 4 that can be very misleading – the dichroism can flip sign between both phases of matter, giving an appearance of a topological signature. This observable is summarized in Figs. 4(b,d), showing sign-flipping harmonics in blue, and sign-maintaining harmonics in red. At first glance, the effect seems reminiscent to the predicted helicity flip [61–63], though the underlying mechanism that determines the sign of the elliptical dichroism is completely different, and is connected with the particular shape of the anomalous ellipticity dependence for each harmonic. Essentially, the sign is determined by whether a certain harmonic tends to maximize for right- or left-elliptical polarizations at a given non-high-symmetry-axis. Remarkably, even though the anomalous ellipticity dependence has very different functional structure between phases, the resulting dichroism is quite similar up to potential sign flips (Figs. 4(a,c), and full data in Appendix D), adding to the misleading nature of the effect. However, further inspection shows that only about half of the observed harmonics flip sign in BiH. Thus, there is almost a uniform distribution of the number of sign-flipping and sign-maintaining harmonics, and there is no collective effect across all harmonics as one would expect from a strong topological signature. Analyzing the harmonic indices, we could not find a clear sequence rule that determines which harmonics flip sign or not (e.g. *6n±1*, *12n±1*, etc.), even using Wolfram Mathematica's 'FindSequenceFunction' functionality [90]. This further suggests a band-structure based mechanism in the non-perturbative optical regime (which would likely change for other driving powers and wavelengths). Furthermore, very few harmonics flip sign in $Na_3Bi$, following trends from other potential topological signatures we considered.

It is crucial to elucidate the microscopic origin of this effect, since if one were to perform an experimental measurement for only few harmonic orders in one material system (say up to the 9$^{th}$), it would seem like a general effect and could be wrongly interpreted. In this context, we recall that the HHG response is necessarily asymmetric when driving with elliptical light along non-high-symmetry-axes. In other words, the dichroism must exist due to the lattice asymmetry, and each harmonic must 'choose' to maximize for either left- or right-elliptical driving. It also must vanish for $\varepsilon=\pm1$. These facts produce roughly similar-looking dichroic structures in the normalized plots in Fig. 4(a,c), even if the anomalous elliptical response is different. Assuming a highly complex and non-perturbative response with both interband and intraband emission channels that interfere, one roughly expects an even distribution of ~50% of harmonics choosing left- and 50% choosing right-polarized helicities (though there is still an inherent connection with the band structure, but one that is not obvious to extract). This is very close to the observed 58%-42% ratio. We would also emphasize that this effect is not observed along high-symmetry axes (as in Fig. 3), further suggesting a non-topological origin.

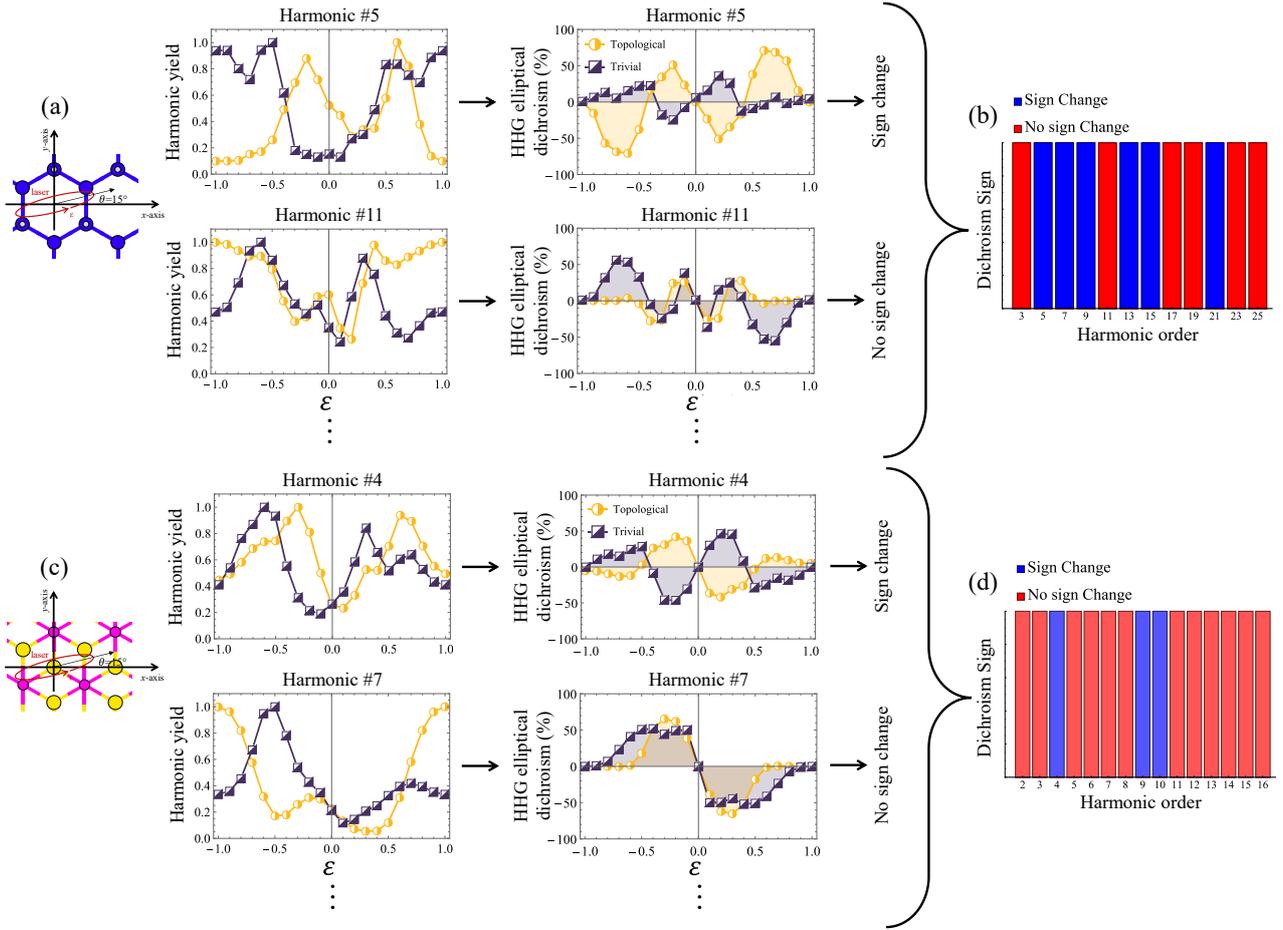

FIG. 4. HHG Elliptical dichroism in monolayers of BiH and Na$_3$Bi, and resulting sign changes in dichroism between the topological phases. (a) Left side: emission from both topologically-trivial BiH without SOC (purple) and the TI phase (yellow), for two select harmonic orders (for full data set see Appendix D, Figs. 17, 18, 21, 22). The yields are normalized to the peak HHG emission for each harmonic order. Right side: resulting elliptical dichroism according to Eq. (9), either producing a dichroism sign change between the phases, or not. (b) Elliptical dichroism sign changes between the topological phases. (c,d) Same as (a,b) but for Na$_3$Bi material system. Plots calculated for similar conditions to Fig. 2 but with elliptical driving and $\theta=15^0$.

### D. Elliptically-polarized driving – Reflection dichroism

To further explore the possible topological character of the elliptical dichroism sign changes we study a related observable that we denote as 'reflection dichroism':

$$RD^n(\theta) = 2\frac{I_\varepsilon^n(\theta) - I_\varepsilon^n(-\theta)}{I_\varepsilon^n(\theta) + I_\varepsilon^n(-\theta)} \tag{11}$$

The dichroic response in Eq. (11) is very similar to that in Eq. (9), but where instead of varying $\varepsilon$ while fixing $\theta$ away from a high-symmetry axis, $\theta$ is varied with $\varepsilon$ fixed away from linear and circular conditions. Moreover, since in BiH the HHG yield is symmetric under $\varepsilon\rightarrow$-$\varepsilon,\theta\rightarrow$-$\theta$, the two equations are mathematically connected. Still, Eqs. (9) and (11) reflect different observables that do not arise from the same measurements – the reflection dichroism contains data for all crystal orientations for a given ellipticity, and not the other way around.

We study this observable by calculating the HHG response *vs.* $\theta$ with a benchmark exemplary ellipticity of $\varepsilon=0.2$. Figure 5 presents the corresponding reflection dichroism for exemplary harmonics (see Appendix D, Figs. 19, 20, 23, 24 for additional data). Overall, the results are analogous to the elliptical dichroism case – the HHG yields show an asymmetric distribution with $\theta$ such that the mirror plane around $\theta=60°$ (apparent under linear driving in Fig. 2) vanishes. The main difference here compared to the elliptical dichroism case is that there is a $60°$ translational symmetry in the HHG yield due to additional crystal symmetries.

The reflection dichroisms in Fig. 5 exhibit five inherent sign changes in BiH for $\theta=0°,30°,60°,90°$, and $120°$ that result from the lattice asymmetry, just as the elliptic dichroism exhibited an inherent sign change at $\varepsilon=0$ and vanished at $\varepsilon=\pm1$. The simple 5-node structure effectively means that all reflection dichroism curves are very similar up to an overall sign, regardless of the phase of matter. Just as in the elliptical dichroism case, this presents a potentially misleading effect where one could wrongfully interpret sign flips between the trivial and topological phases as topological signatures. However, the data for the reflection dichroism further validates that the sign changes are not inherently topological in nature – only roughly two-thirds of the harmonics in BiH exhibit a sign change (Fig. 5(b)), and those harmonic indices are different than those that exhibited sign changes for the elliptical dichroism in Fig. 4(b). If the effect was purely topological, one would expect a consistency. On the other hand, deviations between the sign flipping behavior in the elliptical and reflection dichroisms are clear if considering a structural origin, since the mechanism determining the sign is different in both cases – in Eq. (9) the sign is determined by a tendency to maximize the HHG yield for positive/negative driving helicities away from high-symmetry axes, while in Eq. (11) it is determined by a tendency to maximize the HHG yield for positive/negative driving angles away from the high-symmetry axes. In that respect, the operation of rotating the crystal and changing the driving helicity are not the same.

For $Na_3Bi$, similar conclusions are obtained (see Fig. 5(c,d)), but where the dichroism has slightly reduced symmetry constraints compared to BiH (due to absence of the 6-fold improper-rotation axis). Moreover, just as in the case of the elliptical dichroic response, a smaller number of harmonics change reflection dichroism sign between phases. All of these exhaustive analyses lead us to conclude (at least within the scope of the performed calculations) that the dichroic response in elliptically-driven solid HHG is not uniquely connected with a topological invariant, neither its presence or absence, its magnitude, or its sign.

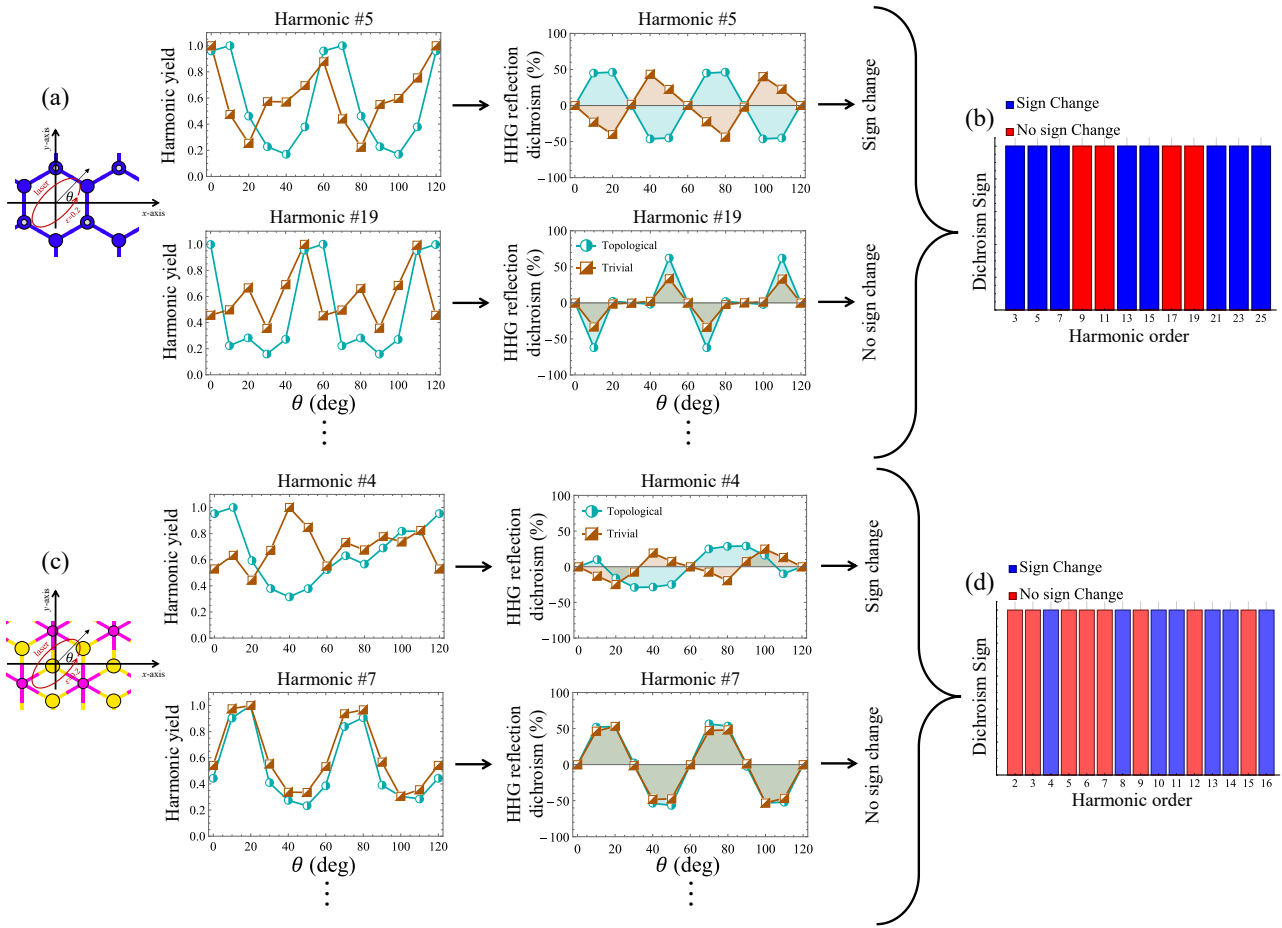

FIG. 5. HHG Reflection dichroism in monolayers of BiH and Na$_3$Bi, and resulting sign changes in dichroism between the topological phases. (a) Left side: emission from topologically-trivial BiH without SOC (brown) and the TI phase (cyan), for two select harmonic orders (for full data set see Appendix D, Figs. 19, 20, 23, 24). The yields are normalized to the peak HHG emission for each harmonic order. Right side: resulting reflection dichroism obtained according to Eq. (11), either producing a dichroism sign change between the phases, or not. (b) Reflection dichroism sign changes between the topological phases. (c,d) Same as (a,b) but for Na$_3$Bi material system. Plots calculated for similar conditions to Fig. 2 but with $\varepsilon=0.2$ and varying elliptical major axis direction.

### E. Temporal characteristics

Lastly, we explore the temporal HHG characteristics by performing a time-frequency analysis (see Appendix A for details). The analysis is performed for the simplest case of linearly-polarized driving along the *x*-axis. Notably, recent work predicted that the harmonic emission time should shift by exactly half a cycle of the driving period (i.e. by $\pi/\omega$) across the topological phase transition collectively for all harmonic orders [71]. This result was obtained for a model Hamiltonian, and was predicted to be valid for all topological systems that lack inversion symmetry (as in Na$_3$Bi). We test it here for both systems, with and without inversion symmetry. Figure 6 shows that there is indeed some temporal shifting in the harmonic emission between topological phases in both materials. However, it is much smaller than a half cycle, and is harmonic order-dependent. Low-order perturbative harmonics shift by about a quarter of a cycle ($\sim 0.5\pi/\omega$) in Na$_3$Bi, and about a tenth of a cycle ($\sim 0.2\pi/\omega$) in BiH. Higher harmonics above the gap (and in the first plateau) shift negligibly in both systems. It is generally difficult to identify delays in the plateau region between phases, since the time-frequency characteristics of the emission change (which is expected as the two phases have different bands and complex electronic structures). Moreover, it was shown that even in the absence of topology, SOC can also introduce HHG time-delays [84]. Therefore, HHG time delays are not unique or universal, are extremely difficult to measure, and highly sensitive to the nature of the bands in the system, making them a poor topological fingerprint.

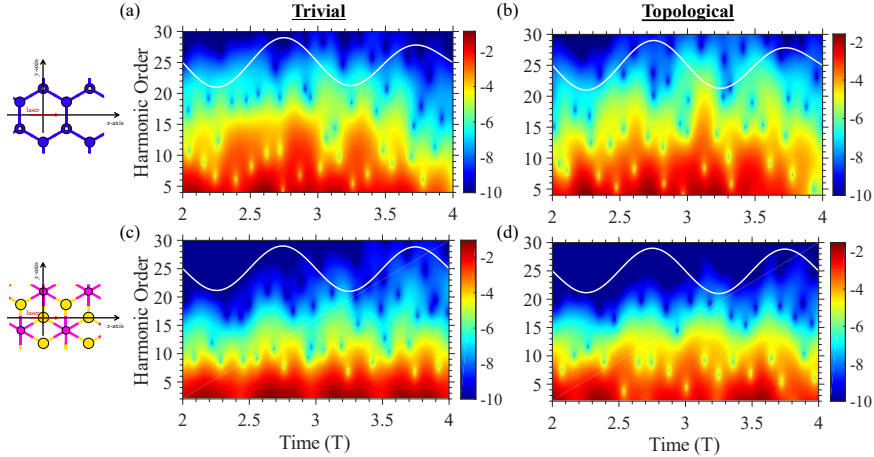

FIG. 6. Time-frequency analysis of HHG emission from monolayers of BiH and Na$_3$Bi for trivial and topological phases. (a) HHG emission *vs.* time (in units of the driving temporal period, T=2π/ω) in BiH for trivial phase. (b) Same as (a) but for the topological phase with SOC. (c) Same as (a) for Na$_3$Bi. (d) Same as (b) for Na$_3$Bi. Calculations performed in similar conditions to Fig. 2 with driving along the *x*-axis. The white line indicates the laser vector potential.

## IV. CONCLUSIONS AND DISCUSSION

To summarize, we presented a thorough *ab-initio* study of HHG from 2D TI in both the quantum spin-Hall and anomalous-Hall phases. Our results compared various HHG-related experimentally-accessible observables of trivial and topological phases of matter. The spectra showed some new potentially promising features that might be utilized for HHS of topology, such as an angular dependence of the HHG yield driven by linearly- or elliptically-polarized lasers, especially in the component parallel to the driving axis and in higher harmonic orders. Nevertheless, our results stress that caution must be taken before interpreting such features as 'topological' in origin, because effects that arise from topological invariants are inherently convoluted with non-topological contributions coming from the band structure, spin-orbit coupling effects, Berry curvatures, and population imbalance. These non-topological contributions dominated the characteristic HHG emission in all examined spectral features and conditions that we analyzed.

We also explicitly showed that some previously employed measures for HHS of topology are not universal and require careful interpretation. This was argued by ruling out the appearance of such signatures in two families of TI in 2D, providing counter examples for their universality and uniqueness. Specifically, the transverse HHG emission component that was predicted to cause a helicity flip in the Haldane model does not exist in the quantum spin-Hall TI, and thus cannot map its topological phase diagram. Moreover, even though it exists in the quantum anomalous-Hall TI, it does not map the phase diagram. The anomalous ellipticity dependence of the HHG yield can also be a problematic topological signature, since we found it arises both in the gapped bulk of a topological phase and in topologically-trivial states, and strongly depends on the driving conditions and material system. Elliptic dichroic responses appeared with similar magnitudes in both trivial and topological phases, indicating that their presence or absence is not a unique topological feature. We thoroughly analyzed the sign of the elliptical dichroism (and reflection dichroism) since at first glance they appeared to be strong topological indicators. However, our detailed analysis revealed that such observables are very misleading, and inherently tied with the crystal and electronic structures, rather than with the Chern number and/or other topological characteristics. Lastly, we showed that time-frequency characteristics of HHG are generally a poor and non-universal topological indicator.

Before ending, we would like to stress that we do not rule out the existence of topological signatures in HHG. Indeed, we expect that the Chern numbers (or other invariants) do get imprinted onto the response. It is however in our opinion much more likely that the topological fingerprints are more subtle and weak compared to the main contributors to the solid HHG response, which are the band structure, lattice symmetry (and asymmetry), and chemical nature of the orbitals. Our lack of success to faithfully extract any signatures of topology from spectra, despite great attempts, is evidence of this difficulty. As a result, universally applicable signatures of topology in HHG seem highly unlikely at this stage. Looking forward, our work should motivate research for formulating quantitative ways to extract topological features from nonlinear spectroscopies in an unambiguous manner, and trigger further studies for developing nonlinear optical spectroscopies in solids. It also raises important questions regarding potential utilization of topology for applications beyond HHG, e.g. in strongly-driven and out-of-equilibrium systems where topology seems to not be the dominant contributor to electronic responses. As a potential outlook, we expect, and hope, that methodologies that perform multi-dimensional HHG spectroscopy with multiple degrees of freedom (e.g. employing CEP-sensitivity [34,67,91], poly-chromatic driving [13,26,29,92–94], pump-probe methodologies [33,95–97], etc.), and perhaps even employing big data techniques and machine learning [98,99], or combinations of multiple spectroscopic probes, might have enhanced topological sensitivity.

## ACKNOWLDENGEMENTS

This work was supported by the Cluster of Excellence Advanced Imaging of Matter (AIM), Grupos Consolidados (IT1249-19), SFB925, "Light induced dynamics and control of correlated quantum systems", has received funding from the European Union's Horizon 2020 research and innovation programme under the Marie Skłodowska-Curie grant agreement No 860553, and NextGenerationEU MUR D.M. 737/2021 "Materials Manipulation with Light". The Flatiron Institute is a division of the Simons Foundation. O.N. gratefully acknowledges the generous support of a Schmidt Science Fellowship.

## APPENDIX A: NUMERICAL DETAILS

We report here on additional technical details for numerical calculations, starting with details of the GS calculations. All DFT calculations were performed with Octopus code [78–80]. The KS equations were discretized on a Cartesian grid of the shape of the primitive lattice cells with a grid spacing of 0.39 Bohr for BiH, and 0.36 Bohr for Na$_3$Bi. Lattice parameters for BiH were taken from ref. [74] ($a=b$=5.53Å, and a Bi-H distance of 1.82Å), and from ref. [76] for Na$_3$Bi ($a=b$=5.448Å, $u$=0.417). The $z$-axis was described with non-periodic boundaries with a total length of 60 Bohr. We employed a Γ-centered 24×24 $k$-grid for BiH including SOC and all Na$_3$Bi systems, and a 30×30 $k$-grid for BiH without SOC (which required a denser sampling around the Dirac cones at K and K'). Deep core states were replaced by Hartwigsen-Goedecker-Hutter (HGH) norm-conserving pseudopotentials [100].

Time propagations employed a time-step of 4.83 attoseconds, and the propagator was represented by a Lanczos expansion without any assumed $k$-point symmetries. Absorbing boundaries were employed through a CAP along the $z$-axis with a width of 15 Bohr and a maximal magnitude of 1 a.u with a sinusoidal shape [101].

The envelope function of the employed laser pulse, $f(t)$, was taken to be of the following 'super-sine' form [82]:

$$f(t) = \left(sin\left(\pi \frac{t}{T_p}\right)\right)^{\left(\frac{\left|\pi\left(\frac{t}{T_p}-\frac{1}{2}\right)\right|}{w}\right)} \tag{12}$$

with $w$=0.75, $T_p$ the duration of the laser pulse which was taken as $T_p$=5$T$ for BiH calculations (~29.3 femtoseconds full-width-half-max (FWHM) for 3000nm light), and $T_p$=6$T$ for Na$_3$Bi calculations (which required slightly longer pulses to avoid CEP issues due to the broken inversion symmetry), where $T$ is a single cycle of the carrier frequency. This form is roughly equivalent to a super-gaussian pulse, but where the field starts and ends exactly at zero amplitude, which is numerically more convenient.

The calculation of the normalized deviation in HHG yield *vs.* driving angle in Fig. 2(b,e) was performed by subtracting the calculated integrated HHG yield of the $n$'th harmonic from the topological phase, $I_n^{topo}(\theta)$, and the trivial phase, $I_n^{triv}(\theta)$. The subtracted normalized yield was integrated over the driving angle to define the following target function: $H_n(\chi) = \int d\theta \left( \frac{I_n^{topo}(\theta)}{\max_\theta\{I_n^{topo}(\theta)\}} - \chi \frac{I_n^{triv}(\theta)}{\max_\theta\{I_n^{triv}(\theta)\}} \right)$, where $\chi$ is a scalar that defines the relative intensity of the emission between phases. $H_n(\chi)$ was minimized for every separate harmonic order with respect to $\chi$ through a least-squares algorithm, which provides a measure for the deviation in the general behavior of the HHG yield between the phases *vs.* $\theta$ (i.e. accounting only for the shape of the curve, and ignoring its overall power). We repeated this process for both material systems, and also for separating the HHG yield associated with both transverse and parallel emission components.

The time-frequency analysis in Fig. 6 is performed with a Gabor analysis [102] of the total current with a gaussian window of temporal width of a third of a cycle of the driving period.

### APPENDIX B: RADIAL HHG EMISSION SPECTROGRAMS

We present here results complementary to those in the main text that presented the integrated HHG yield per harmonic order for certain harmonics. Here we present the full unprocessed data of HHG spectra driven by linearly-polarized light, separated to parallel components (Fig. 7), and transverse components (Fig. 8), in both materials and phases. The integrated yields per harmonic order are presented in Appendix D.

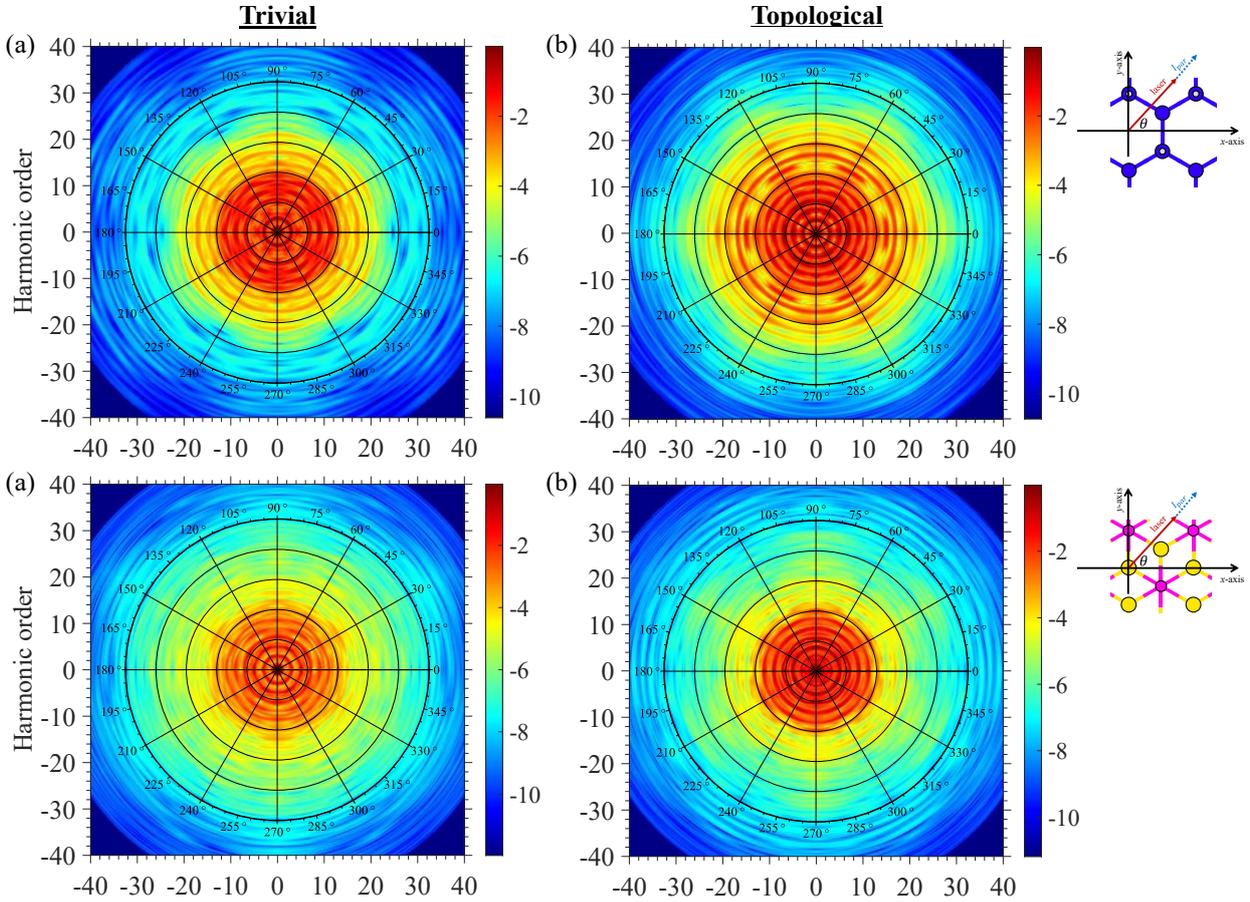

FIG. 7. Parallel HHG spectra *vs.* crystal orientation from BiH and Na$_3$Bi driven by a linearly-polarized laser. (a,b) Results from both phases of BiH. (c,d) Results from both phases of Na$_3$Bi. The calculation is performed in the same conditions as Fig. 2.

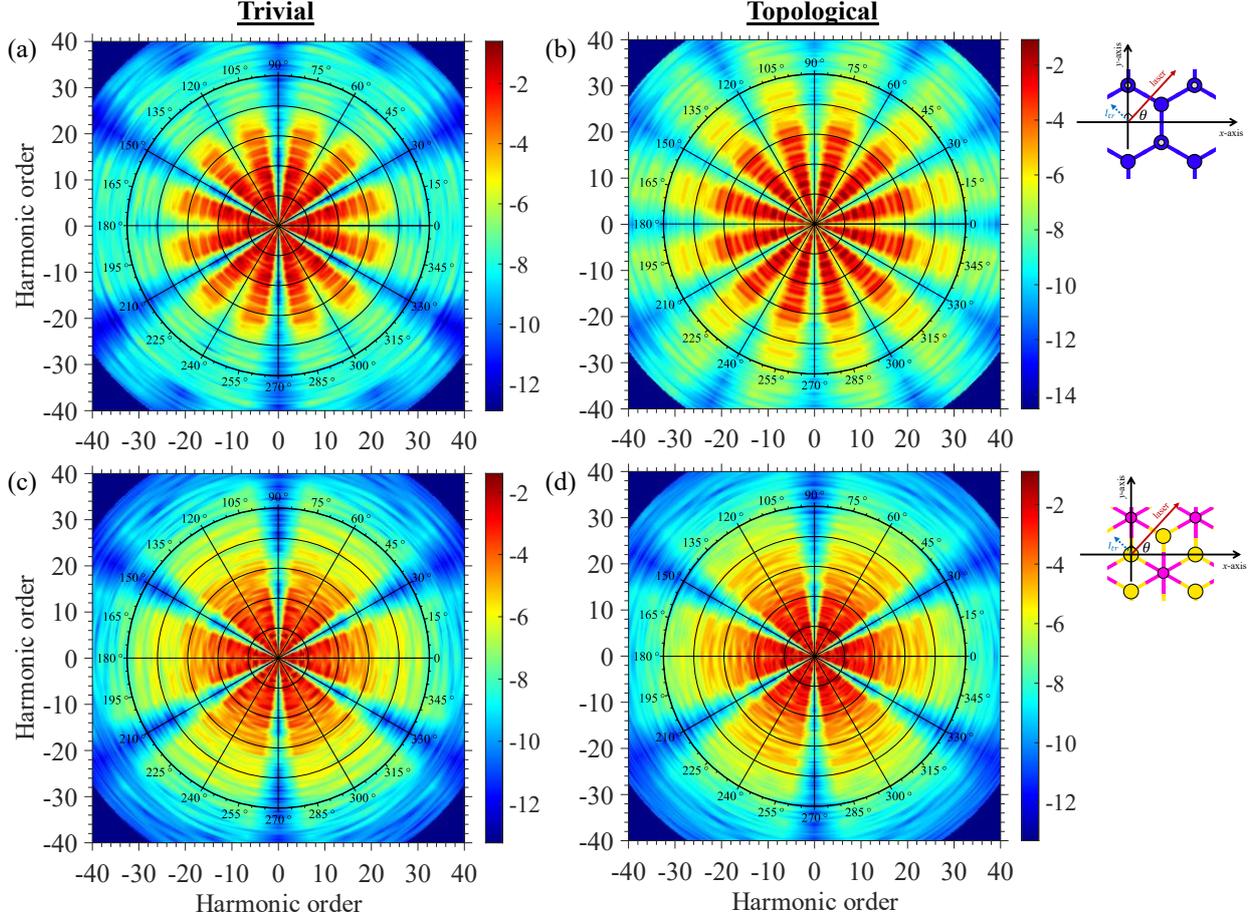

FIG. 8. Transverse HHG spectra *vs.* crystal orientation from BiH and Na$_3$Bi driven by a linearly-polarized laser. (a,b) Results from both phases of BiH. (c,d) Results from both phases of Na$_3$Bi. The calculation is performed in the same conditions as Fig. 2. The transverse components reveal the point group symmetries of the materials, which slightly differ.

## APPENDIX C: HHG EMISSION FROM INDIVIDUAL SPIN CHANNELS

Figure 9 shows the transverse HHG emission components for the linear-driving case *vs.* $\theta$ for the topological phases of BiH and Na$_3$Bi (with SOC). From Fig. 9 we see that the HHG emission from each separate spin channel allows nonzero transverse response even along high-symmetry crystal axes. For BiH the overall transverse response vanishes when both spin channel contributions are coherently summed (because the transverse components have opposite signs in the quantum spin-Hall phase). In Na$_3$Bi each spin channel emits a chiral-like patten that lacks mirror symmetries. The summation of both channels remains nonzero due to the inversion symmetry breaking in the crystal structure, allowing elliptical harmonics to be emitted, and transverse HHG emission even when driven along the *x*-axis.

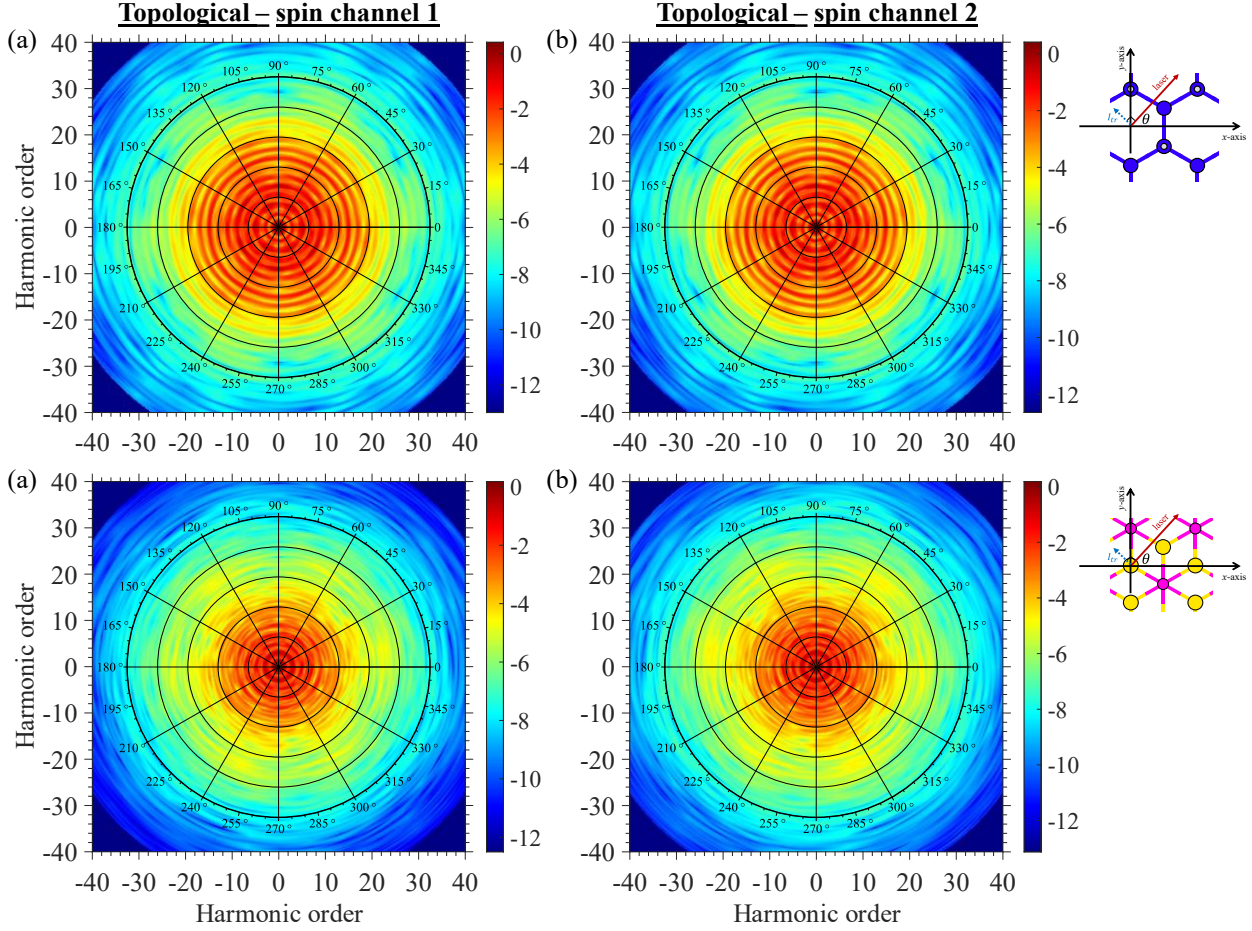

FIG. 9. Transverse HHG spectra *vs.* crystal orientation from monolayers BiH and Na$_3$Bi driven by a linearly-polarized laser for the topological phase (with SOC). (a,b) Same as (a,b) in Fig. 8, but for each individual spin channel components. (c,d) Same as (a,b) but in Na$_3$Bi.

# APPENDIX D: HARMONIC-RESOLVED DATA

We present here all complementary data that supplements and supports conclusions drawn in the main text. Specifically, the HHG integrated yield resolved per harmonic order in all different driving conditions discussed, compared between topological and trivial phases for both material systems. This includes: linear driving with varying laser polarization angle (Figs. 10-13), elliptical driving with varying ellipticity and main elliptical axis (Figs. 14-16, 17, 19, 21, 23), and the resulting elliptical and reflection dichroism data (Figs. 18, 20, 22, 24). Each figure caption describes the data set and has an inset that graphically explains which material system and laser driving condition is considered.

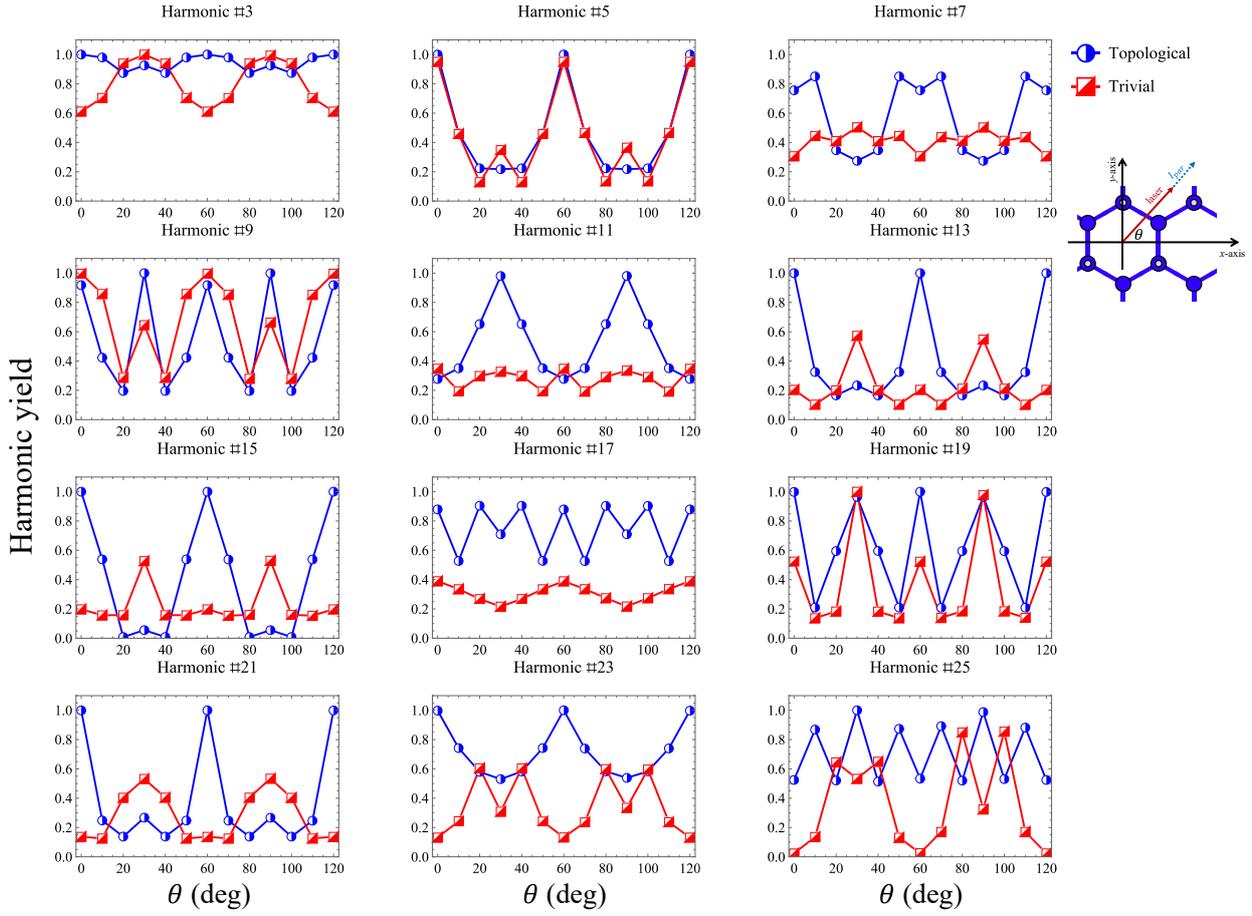

FIG. 10. Parallel HHG spectra components (integrated and normalized per harmonic order) *vs.* crystal orientation from monolayer BiH driven by a linearly-polarized laser for the topological phase with SOC (blue) and trivial phase without SOC (red). Calculated for the same driving conditions as in Fig. 2, and complementing the two select harmonics presented in Fig. 2(a). Fig. 2(b,c) is derived from this data.

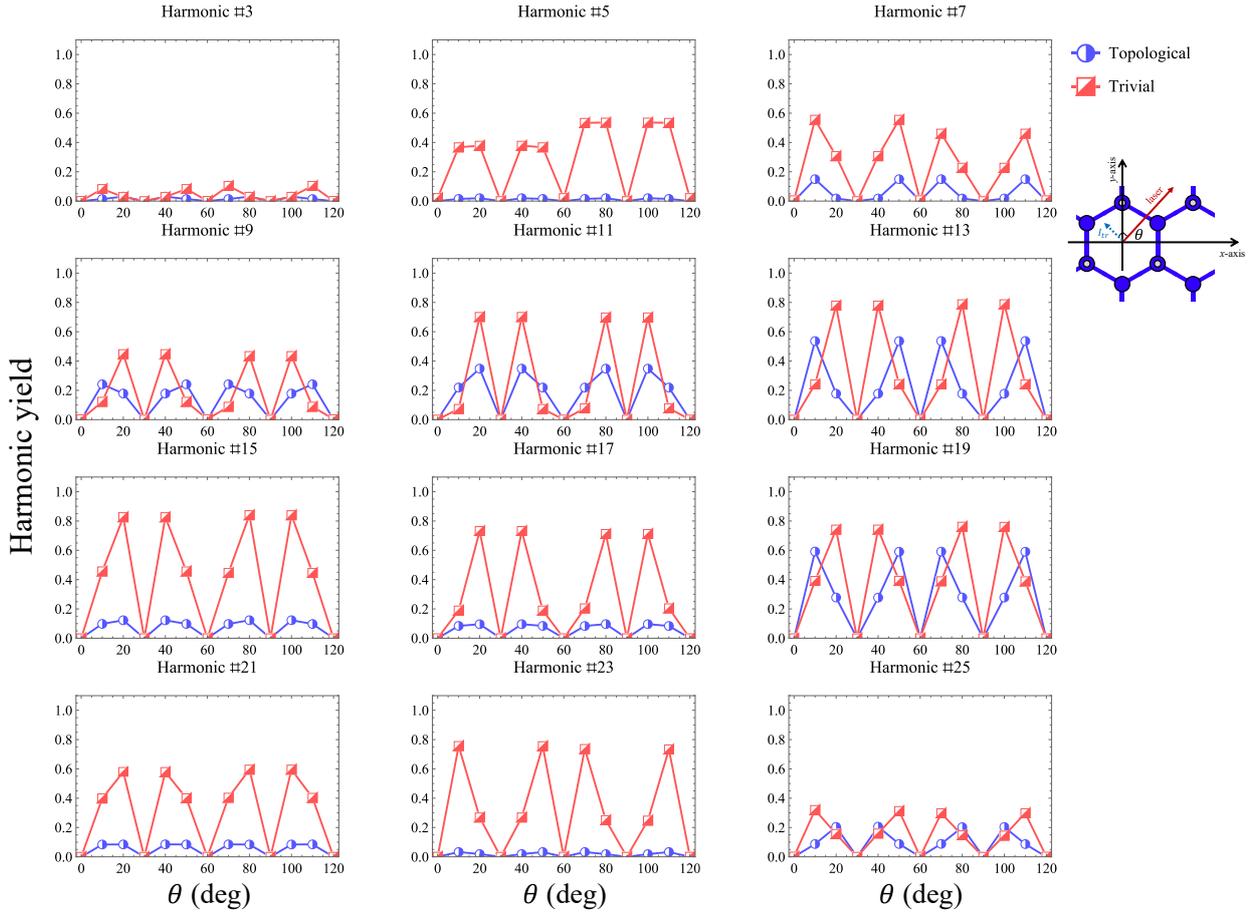

FIG. 11. Transverse HHG spectra components (integrated and normalized per harmonic order) *vs.* crystal orientation from monolayer BiH driven by a linearly-polarized laser for the topological phase with SOC (blue) and trivial phase without SOC (red). Calculated for the same driving conditions as in Fig. 2, and complementing the two select harmonics presented in Fig. 2(a). Fig. 2(b,c) is derived from this data.

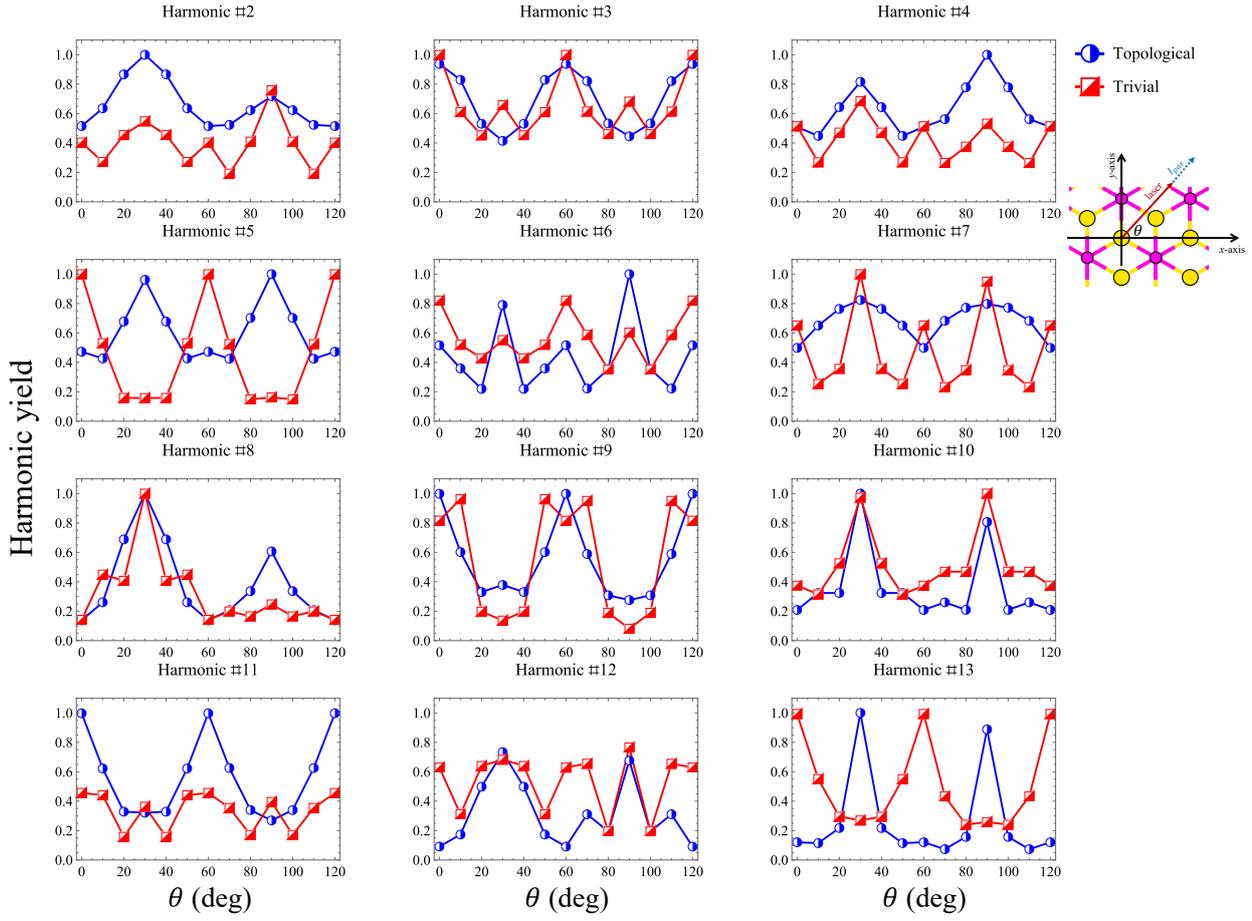

FIG. 12. Parallel HHG spectra components (integrated and normalized per harmonic order) *vs.* crystal orientation from monolayer Na$_3$Bi driven by a linearly-polarized laser for the topological phase with SOC (blue) and trivial phase without SOC (red). Calculated for the same driving conditions as in Fig. 2, and complementing the two select harmonics presented in Fig. 2(d). Fig. 2(e,f) is derived from this data.

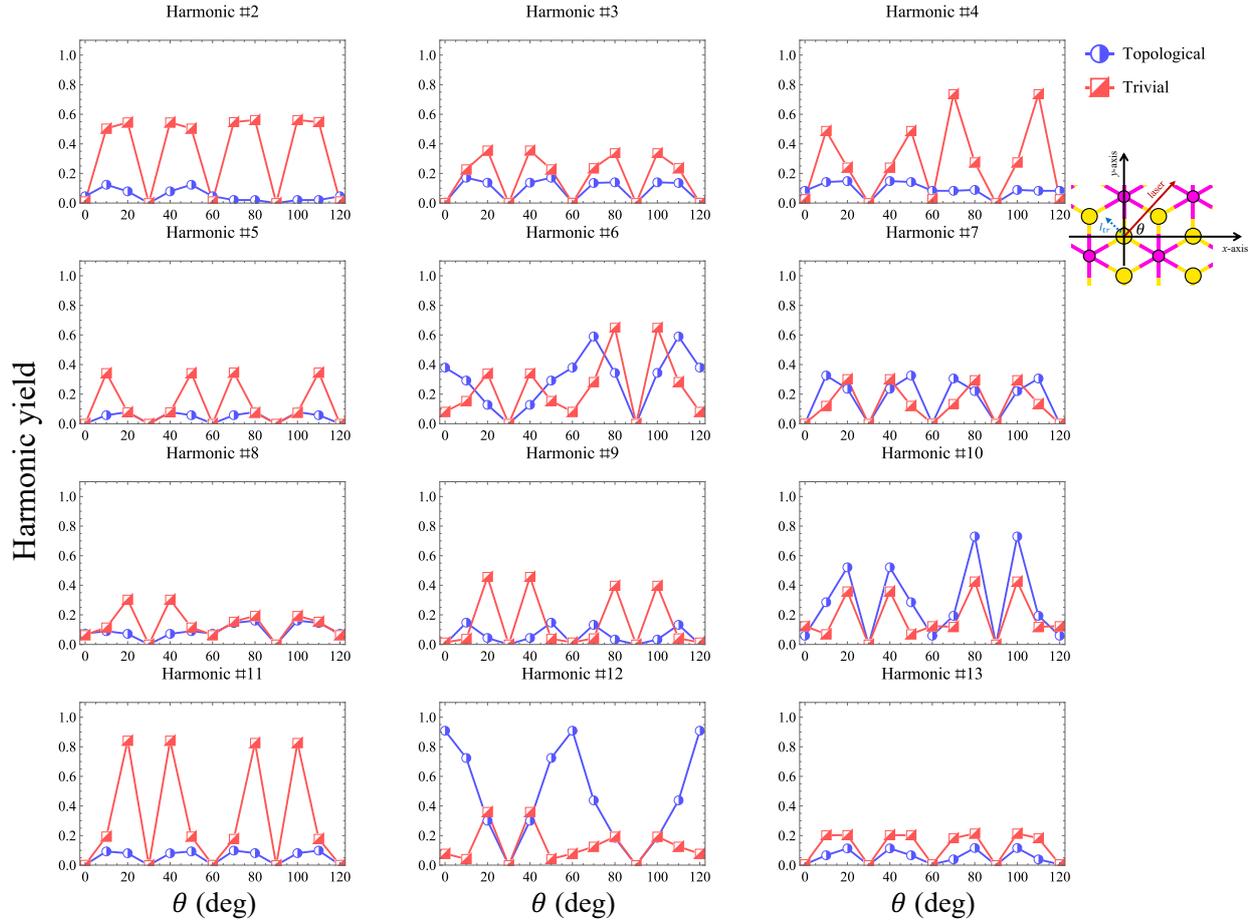

FIG. 13. Transverse HHG spectra components (integrated and normalized per harmonic order) *vs.* crystal orientation from monolayer Na$_3$Bi driven by a linearly-polarized laser for the topological phase with SOC (blue) and trivial phase without SOC (red). Calculated for the same driving conditions as in Fig. 2, and complementing the two select harmonics presented in Fig. 2(d). Fig. 2(e,f) is derived from this data.

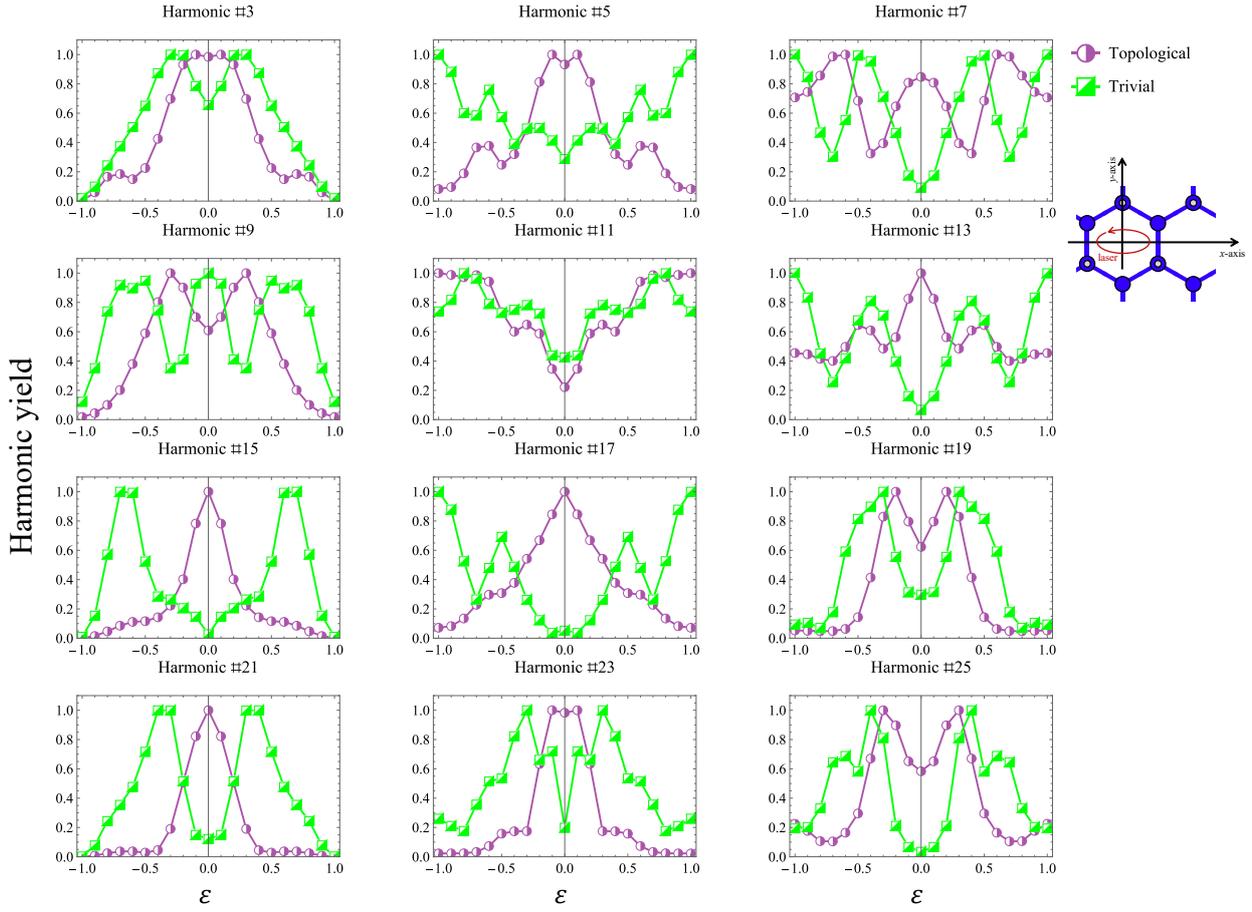

FIG. 14. HHG yield (integrated and normalized per harmonic order) *vs.* laser ellipticity from monolayer BiH for the topological phase with SOC (purple) and trivial phase without SOC (green). Calculated for the same driving conditions as in Fig. 3, and complementing the three select harmonics presented in Fig. 3(a).

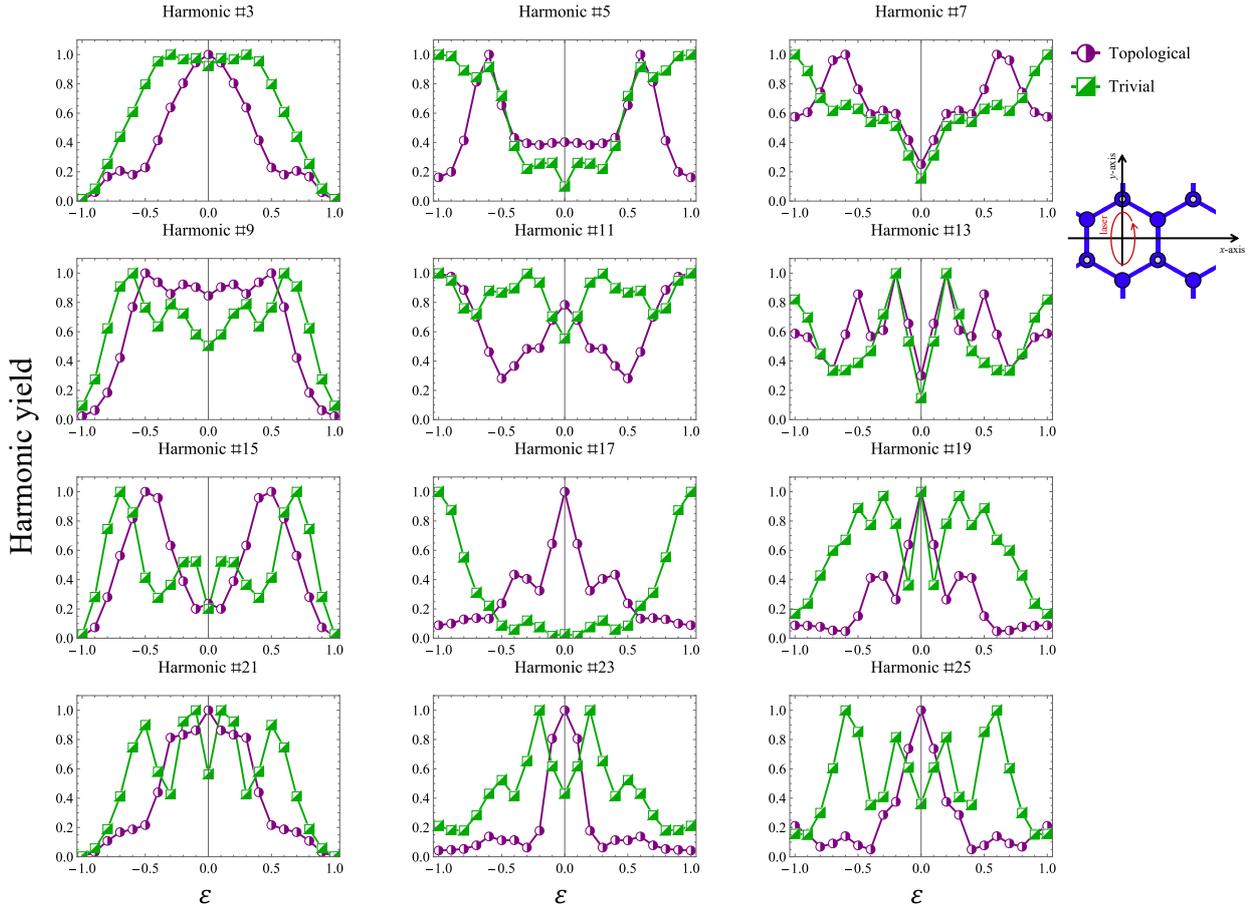

FIG. 15. HHG yield (integrated and normalized per harmonic order) *vs.* laser ellipticity from monolayer BiH for the topological phase with SOC (purple) and trivial phase without SOC (green). Calculated for the same driving conditions as in Fig. 3, and complementing the three select harmonics presented in Fig. 3(b).

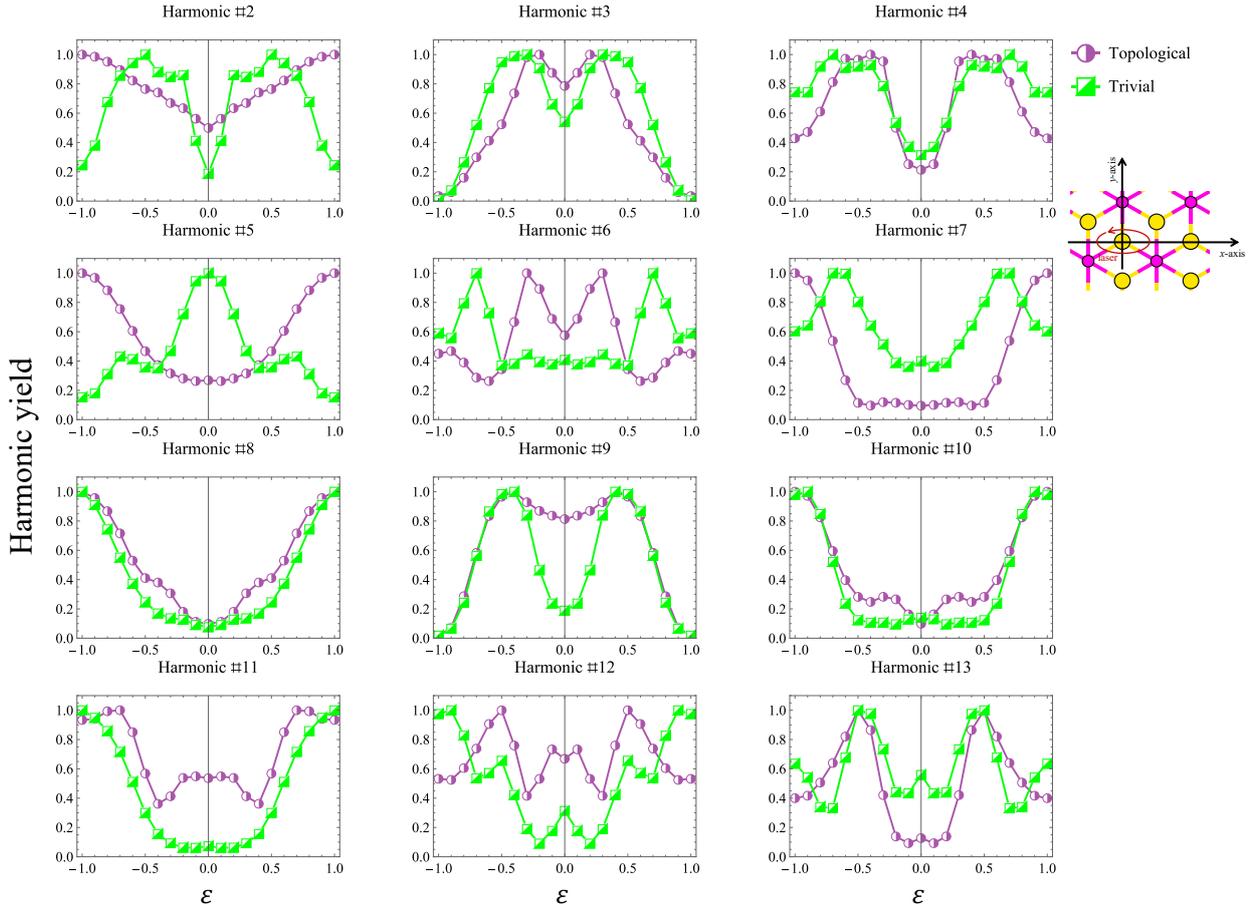

FIG. 16. HHG yield (integrated and normalized per harmonic order) *vs.* laser ellipticity from monolayer Na$_3$Bi for the topological phase with SOC (purple) and trivial phase without SOC (green). Calculated for the same driving conditions as in Fig. 3, and complementing the three select harmonics presented in Fig. 3(c).

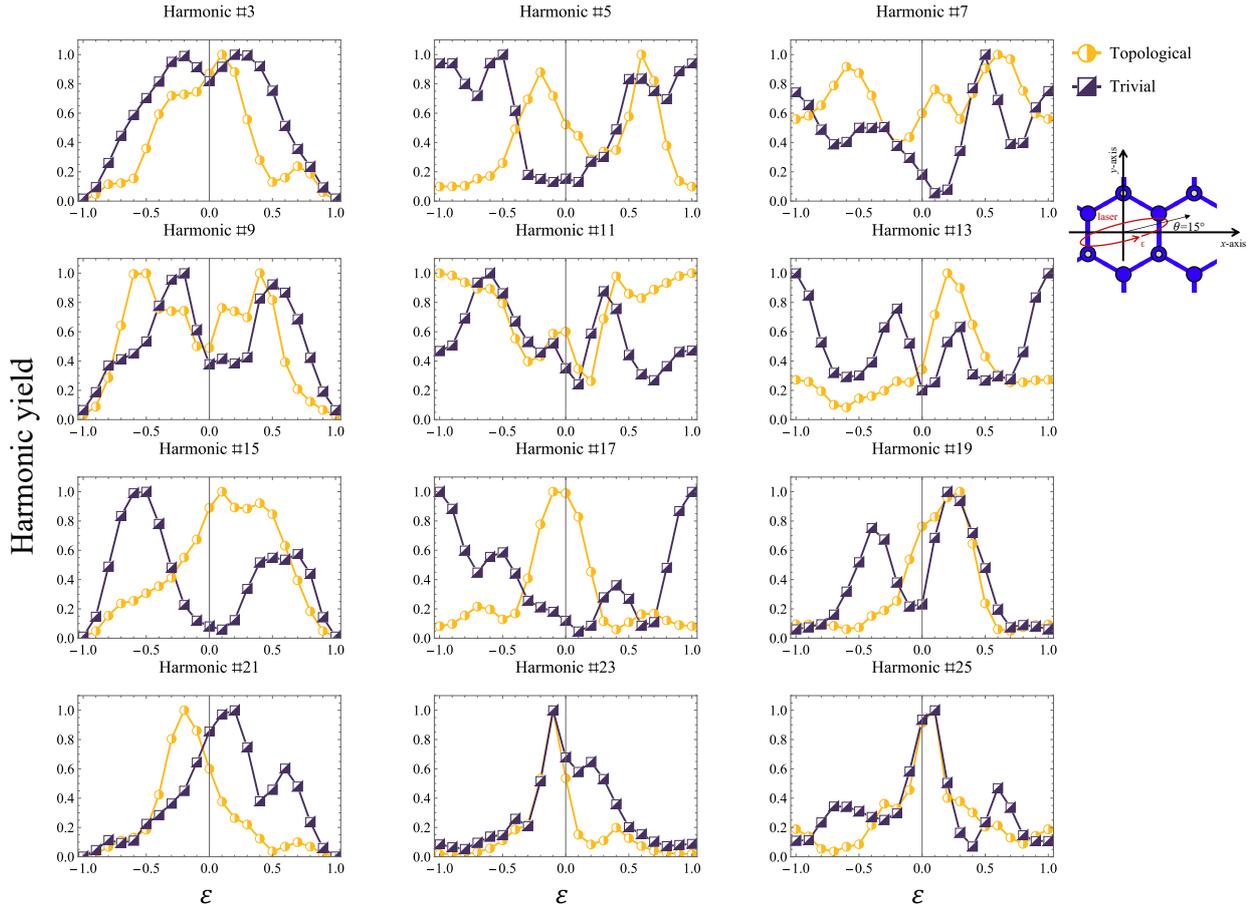

FIG. 17. HHG yield (integrated and normalized per harmonic order) *vs.* laser ellipticity from monolayer BiH for the topological phase with SOC (yellow) and trivial phase without SOC (purple). Calculated for the same driving conditions as in Fig. 4, and complementing the two select harmonics presented in Fig. 4(a). Fig. 4(b) is derived from this data.

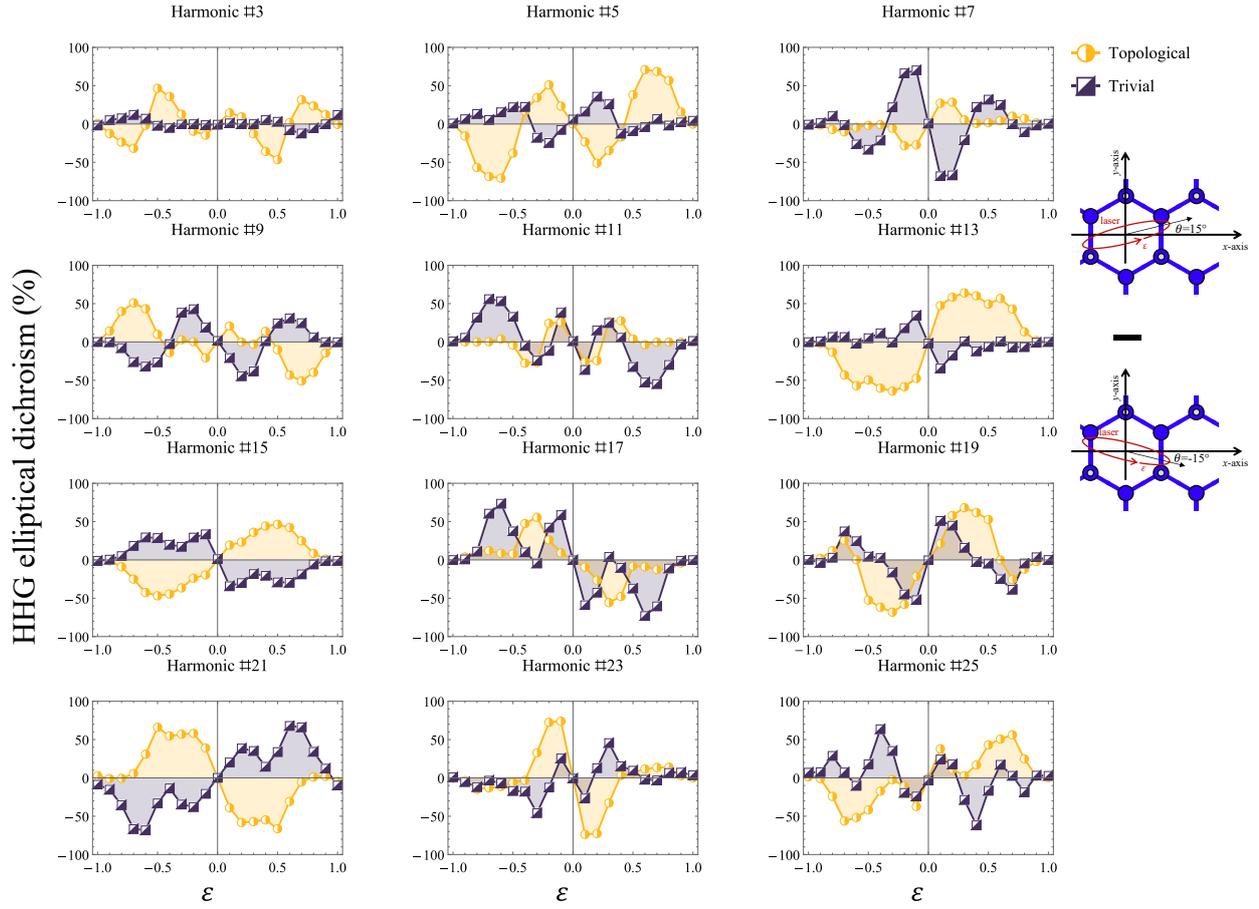

FIG. 18. HHG elliptical dichroism *vs.* laser ellipticity from monolayer BiH for the topological phase with SOC (yellow) and trivial phase without SOC (purple). Calculated for the same driving conditions as in Fig. 4, and complementing the two select harmonics presented in Fig. 4(a). Fig. 4(b) is derived from this data.

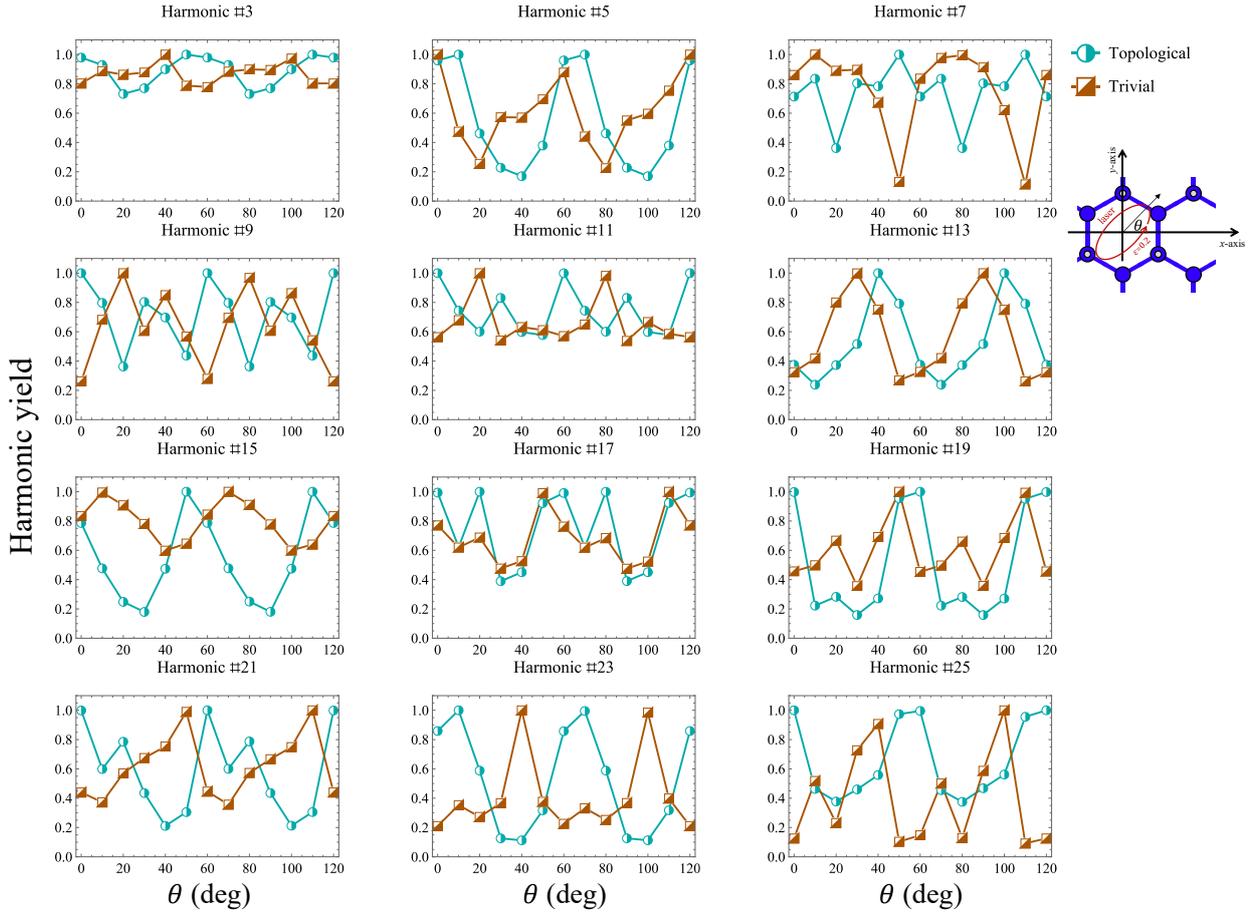

FIG. 19. HHG yield (integrated and normalized per harmonic order) *vs.* laser major elliptical axis angle from monolayer BiH for the topological phase with SOC (cyan) and trivial phase without SOC (brown). Calculated for the same driving conditions as in Fig. 5, and complementing the two select harmonics presented in Fig. 5(a). Fig. 5(b) is derived from this data.

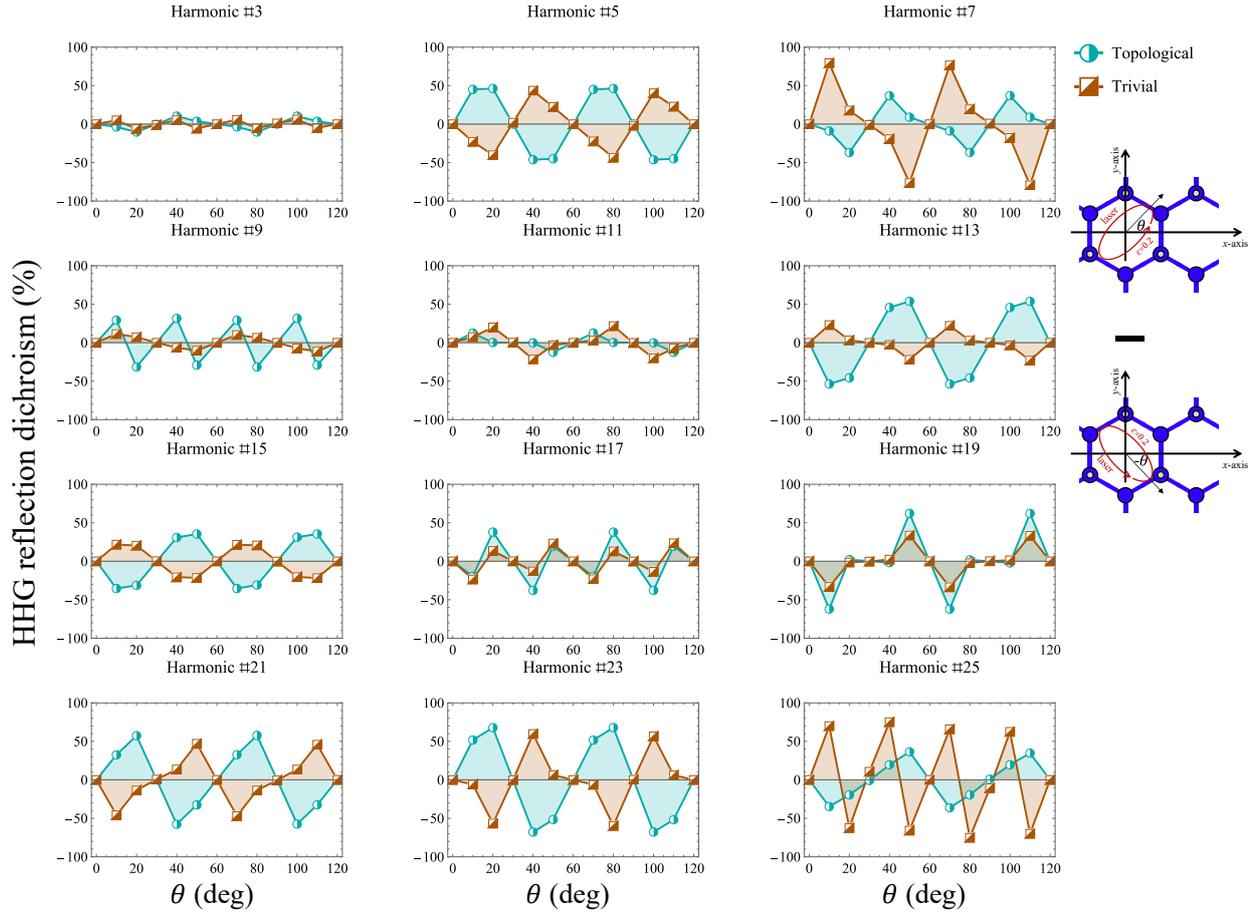

FIG. 20. HHG reflection dichroism *vs.* laser major elliptical axis angle from monolayer BiH for the topological phase with SOC (cyan) and trivial phase without SOC (brown). Calculated for the same driving conditions as in Fig. 5, and complementing the two select harmonics presented in Fig. 5(a). Fig. 5(b) is derived from this data.

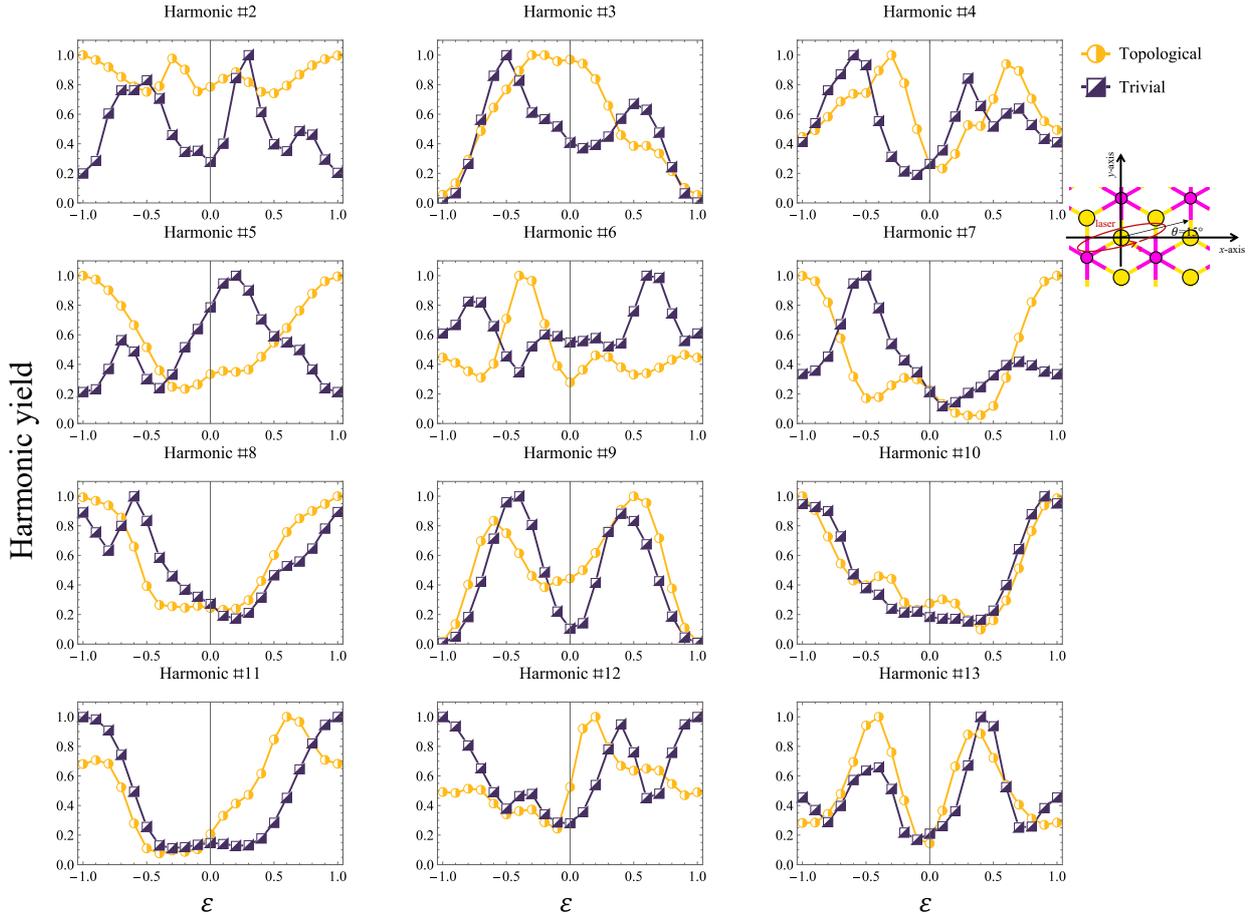

FIG. 21. HHG yield (integrated and normalized per harmonic order) *vs.* laser ellipticity from monolayer Na$_3$Bi for the topological phase with SOC (yellow) and trivial phase without SOC (purple). Calculated for the same driving conditions as in Fig. 4, and complementing the two select harmonics presented in Fig. 4(c). Fig. 4(d) is derived from this data.

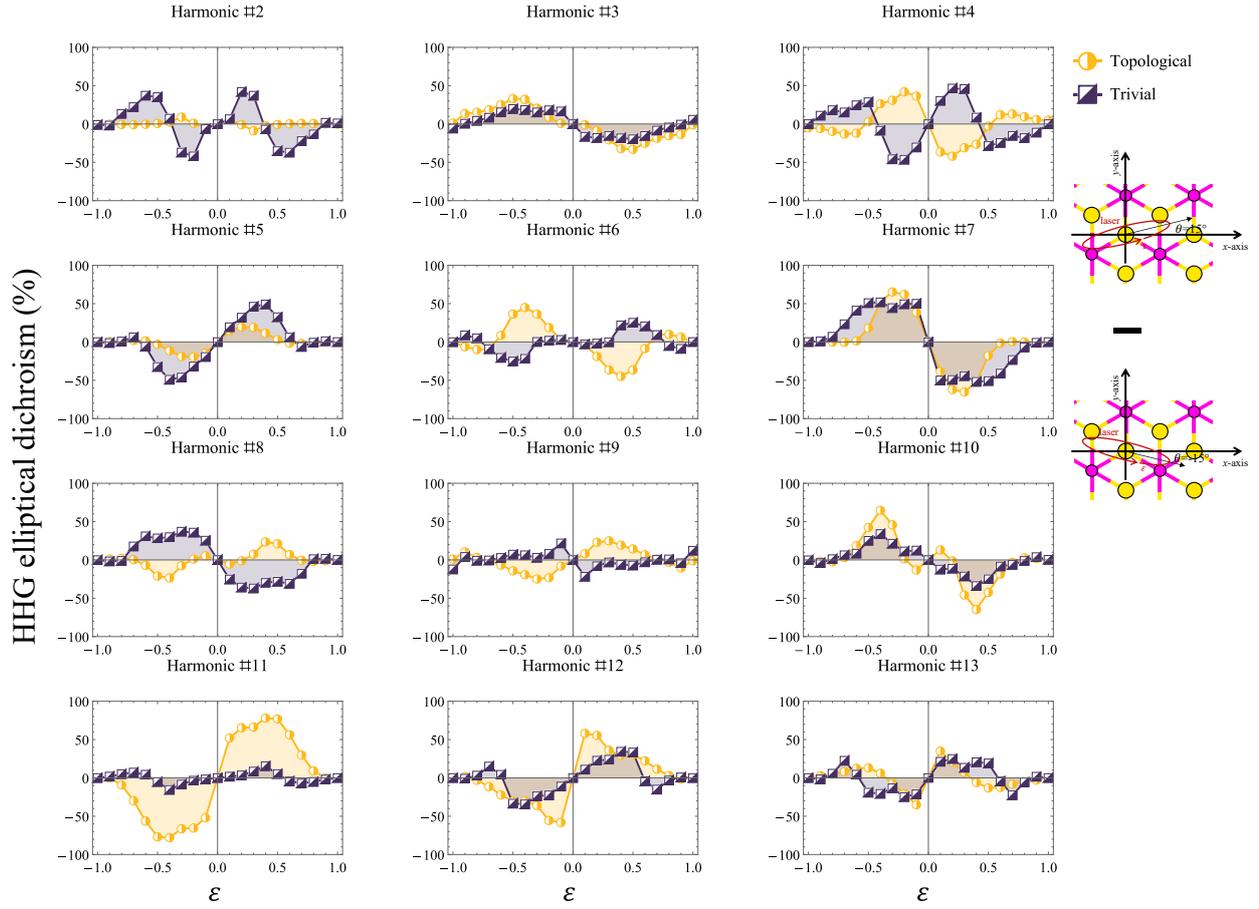

FIG. 22. HHG elliptical dichroism *vs.* laser ellipticity from monolayer Na$_3$Bi for the topological phase with SOC (yellow) and trivial phase without SOC (purple). Calculated for the same driving conditions as in Fig. 4, and complementing the two select harmonics presented in Fig. 4(c). Fig. 4(d) is derived from this data.

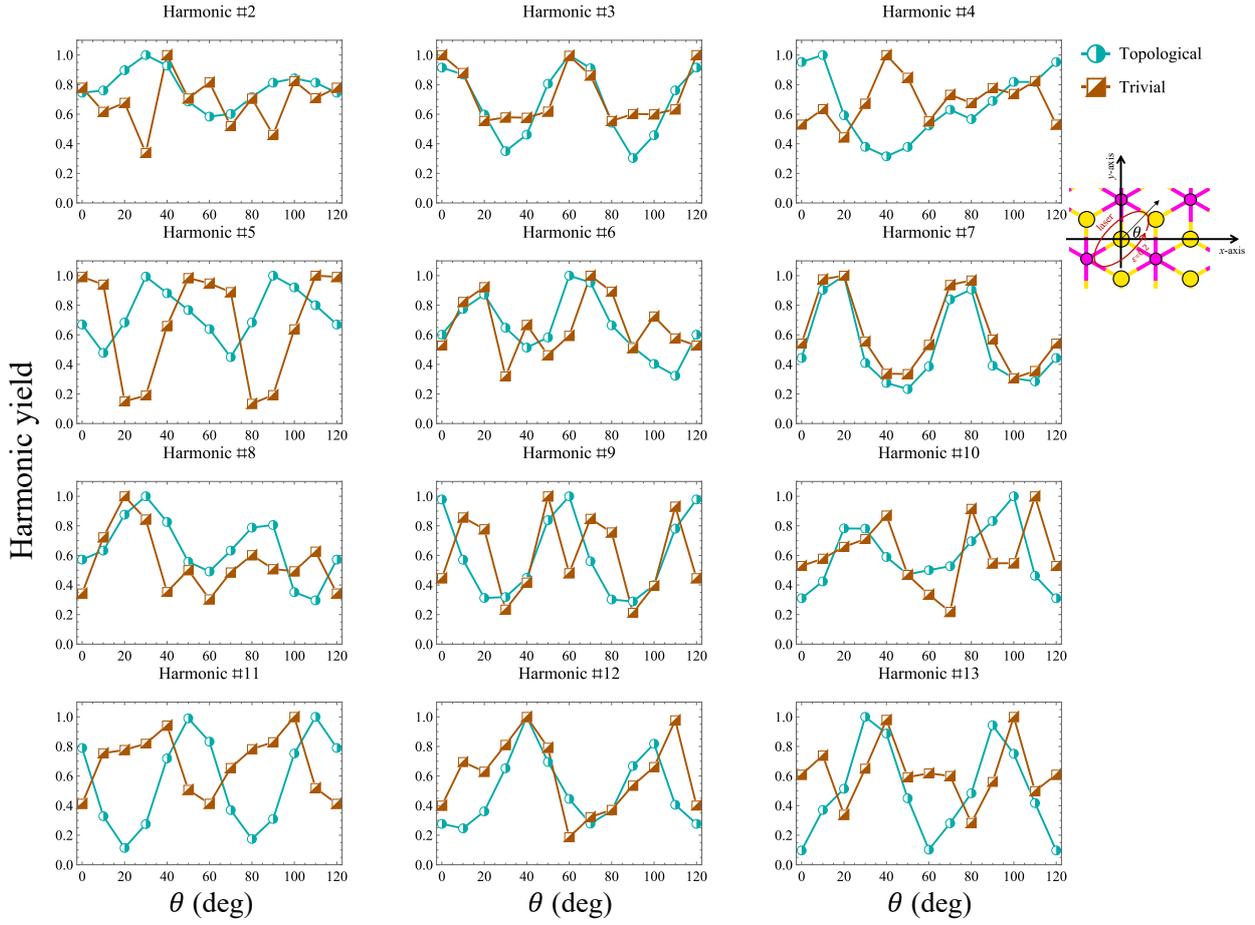

FIG. 23. HHG yield (integrated and normalized per harmonic order) *vs.* laser major elliptical axis angle from monolayer BiH for the topological phase with SOC (cyan) and trivial phase without SOC (brown). Calculated for the same driving conditions as in Fig. 5, and complementing the two select harmonics presented in Fig. 5(c). Fig. 5(d) is derived from this data.

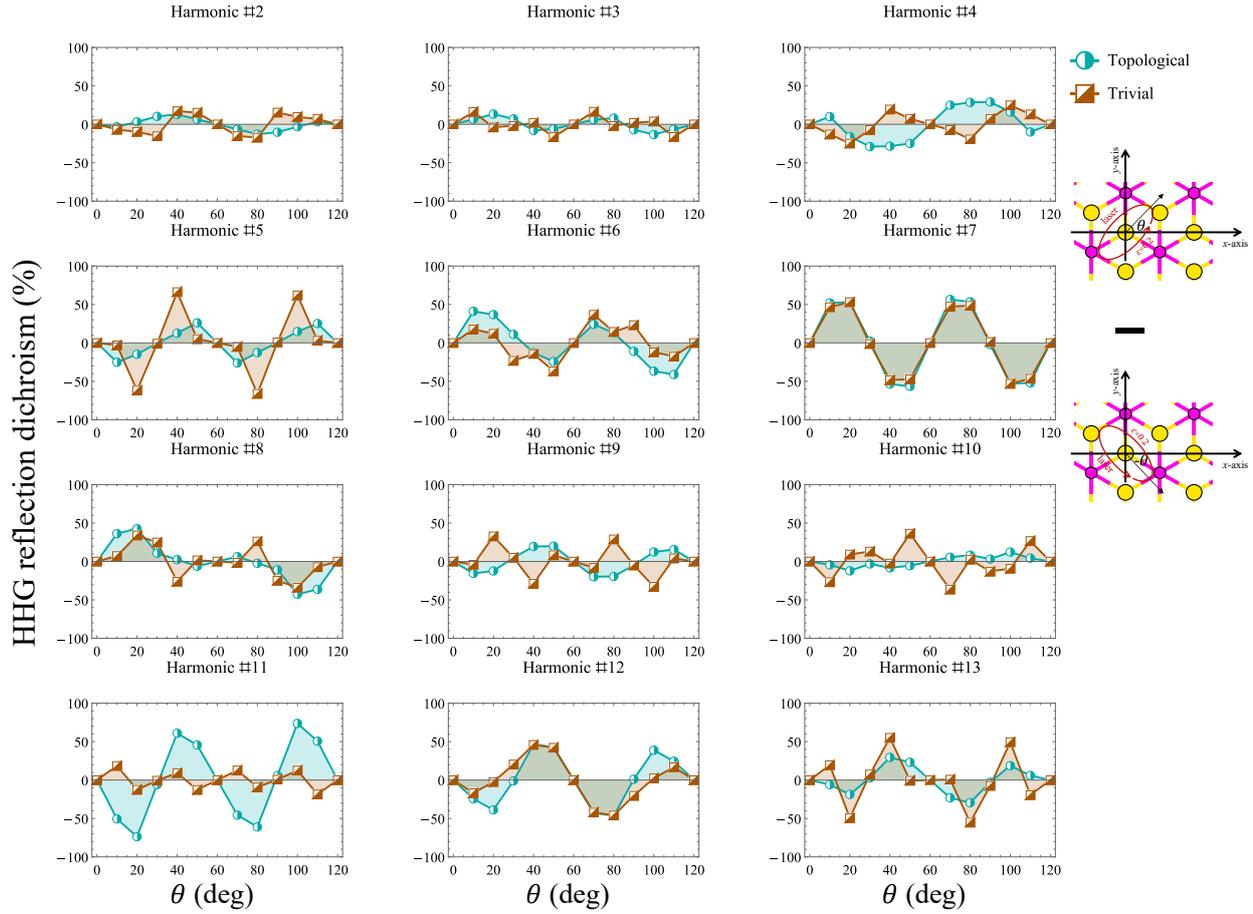

FIG. 24. HHG reflection dichroism *vs.* laser major elliptical axis angle from monolayer $Na_3Bi$ for the topological phase with SOC (cyan) and trivial phase without SOC (brown). Calculated for the same driving conditions as in Fig. 5, and complementing the two select harmonics presented in Fig. 5(c). Fig. 5(d) is derived from this data.